\RequirePackage{fix-cm}
\documentclass[smallextended]{svjour3}       
\smartqed  
\usepackage{graphicx}
\usepackage{amssymb}

\usepackage{bm}

\usepackage{amsmath}

\newcommand{\be}{\begin{equation}}
\newcommand{\ee}{\end{equation}}

\usepackage{epsfig,amsmath,amssymb}

\font\tenscr=rsfs10 scaled1100
\font\sevenscr=rsfs7 
\font\fivescr=rsfs5 
\skewchar\tenscr='177 \skewchar\sevenscr='177 \skewchar\fivescr='177
\newfam\scrfam
\textfont\scrfam=\tenscr
\scriptfont\scrfam=\sevenscr
\scriptscriptfont\scrfam=\fivescr

\def\scri{{\fam\scrfam I}}

\begin{document}


\title{
On the representation theory of the
Bondi$-$Metzner$-$Sachs  group and its variants
in three space$-$time dimensions
}



\author{Evangelos Melas
}


\institute{Evangelos Melas \at
              University of Athens \at
              Department of Economics \at
              Unit of Mathematics and Informatics \at
              Sofokleous 1 \at
              10559 Athens \at
              Greece \\
              Tel.: +030-210-3689403\\
              Fax: +030-210-3689471\\
              \email{emelas@econ.uoa.gr}           
%
}

\date{Received: date / Accepted: date}

\maketitle

\begin{abstract}
The  original  Bondi$-$Metzner$-$Sachs (BMS) group $B$ is the common asymptotic symmetry group of all asymptotically flat Lorentzian radiating 4$-$dim space$-$times. As such, $B$ is  the best candidate for the universal symmetry group of General Relativity (G.R.). In 1973,  with this motivation, P. J. McCarthy classified all relativistic $B$$-$invariant systems in terms of strongly continuous irreducible unitary representations (IRS) of $B$. Here we introduce the analogue $B(2,1)$  of 
the BMS group $B$
in 3 space$-$time dimensions. $B(2,1)$ itself
admits thirty$-$four  analogues both real
in all signatures and in complex space$-$times.
In order to find the IRS of both $B(2,1)$ and its analogues
we need to extend Wigner$-$Mackey's theory of induced
representations. The necessary extension is
described and is
reduced to the
solution of three problems. These  problems are solved in
the case where $B(2,1)$ and its analogues are equipped
with the Hilbert topology. The extended theory is necessary
in order to construct the IRS
of both $B$ and its analogues in any number $d$
of space$-$time dimensions, $d\geq3$, and also in order to construct the IRS
of their supersymmetric counterparts.
We use the extended theory  to obtain the necessary data in order to construct the IRS
of $B(2,1).$
The main results of the representation theory are:
The IRS are induced from ``little groups'' which are compact. The finite  ``little groups'' are cyclic groups of even order.
The inducing construction is exhaustive notwithstanding the fact that $B(2,1)$
is not locally compact in the employed Hilbert topology.

\PACS{04 \and  03 \and 02 \and 11}
\end{abstract}

\section{Introduction}
\label{intro}
The
BMS group $B$ is the common asymptotic group of all
curved real Lorentzian radiating space$-$times which are  asymptotically flat in future null directions \cite{Bondi,Sachs1}, and is the best candidate for the universal symmetry group of G.R..

The BMS group $B$ of general relativity was first
discovered \cite{Bondi,Sachs1} not as a transformation group of
(exact) global diffeomorphisms of a fixed manifold,
but as a pseudo$-$group of local diffeomorphisms
(``asymptotic isometries") of the asymptotic region
of Lorentzian radiating space$-$times which are asymptotically
flat in lightlike future directions.

However, Penrose \cite{Pen1,Pen2,Pen3} showed that, by ``going to infinity", $B$ could be interpreted as an (exact) global transformation group $B \times \Im^{+} \longrightarrow \Im^{+}$ of the ``future null boundary" $\Im^{+}$, or, of the
so called ``future null infinity", of the space$-$times concerned. Furthermore, he gave \cite{Pen1,Pen2}  a geometric structure to $\Im^{+}$, the ``strong conformal geometry" for which the
transformation group is precisely the group of automorphisms
\cite{Pen3}.

In 1939 Wigner laid the foundations
of special relativistic quantum mechanics \cite{Wigner} and relativistic quantum field theory by constructing
the Hilbert space
IRS of the (universal cover) of the Poincare group $P$.

The universal property of $B$ for G.R.
makes it reasonable to attempt
to lay a similarly firm foundation for quantum gravity by following through the analogue of
Wigner's programme with $B$ replacing $P$. Some years ago McCarthy constructed explicitly \cite{mac3,mac8,mac4,mac5,mac6,mac7,Crampin2,Mac}   the IRS of $B$ for exactly this purpose.
This work was based on G.W.Mackey's pioneering work on group representations \cite{Mackey0,Mackey00,Mackey,Mackey1,Simms,Isham}; in particular
it was based on an extension \cite{mac6} of G.W.Mackey's  work to the relevant
infinite dimensional case.


It is difficult to overemphasize the importance of Piard's results \cite{Piard1,Piard2} who soon afterwards proved that \it{all} \normalfont the IRS of $B$,
when this is equipped with the Hilbert topology, are derivable by the inducing construction.
This proves the exhaustivity of McCarthy's list of representations and renders his results even more important.

However, in quantum gravity, complexified or euclidean versions of G.R. are frequently considered and
the question arises: Are there similar symmetry groups for these versions of the theory? McCarthy constructed \cite{mac1}, in abstract form,
all possible analogues of $B$, both real and in any signature, or complex, with all possible notions of asymptotic flatness ``near infinity''. There are, in fact, forty$-$one such groups.
One of them is $B(2,2)$, the asymptotic symmetry group of all curved real ultrayperbolic space$-$times which are asymptotically flat in null directions.

These abstract constructions were given in a quantum setting; the paper \cite{mac1} was concerned with finding the IRS of these groups
in Hilbert spaces (especially for the complexification $ \mathcal C \mathcal B $ of $B$ itself). It has been argued \cite{mac1,Mel1}
that these Hilbert space representations are
related to elementary particles and quantum gravity (via gravitational instantons). The IRS of $B(2,2)$ were constructed in \cite{macMel,Mel5}.
HB, a subgroup of $B(2,2)$ which arises naturally in the construction of the analogues of  $B$ by McCarthy \cite{mac1}, and which, surprisingly, remained unnoticed by him, was introduced in \cite{Melas1}. The representation theory of HB was commenced in \cite{Mel6,Mel7}.

Here,
we initiate                                      this programme for 3$-$dim G.R..
We define the analogue $B(2,1)$ of the Bondi$-$Metzner$-$Sachs (BMS) group
in 3 space$-$time dimensions.
We construct the IRS of
of $B(2,1)$
in the Hilbert topology.
McCarthy and Crampin argued convincingly
in \cite{Crampin2} that
in the case of the original
BMS group $B$, in 4
space$-$time dimensions,
IRS
in the Hilbert topology describe
\it bounded \normalfont sources \cite{Crampin2}, i.e., $B-$elemenatry particles.
Moreover they pointed out that passing from
the coarser Hilbert topology to the finer
nuclear topology the effect is to obtain more
IRS which now also accommodate scattering states of $B-$elemenatry particles.

Their arguments apply also in the case of
$B(2,1)$, in 3
space$-$time dimensions, so we expect IRS of $B(2,1)$
in the Hilbert topology to describe
bounded sources, i.e., $B(2,1)-$elemenatry particles, whereas IRS in the
nuclear topology,
to describe also scattering states of $B(2,1)-$elemenatry particles. Besides
the Hilbert and the nuclear topologies there
is a  wide range
of
``reasonable'' topologies by which we can endow $B(2,1)$ with. These topologies are elaborated in Section \ref{top}.
$B(2,1)$ admits generalizations both real
in all signatures and in complex space$-$times
with all possible notions of asymptotic
flatness ``near infinity''. There are, in fact, thirty$-$five such groups including $B(2,1)$ \cite{Mel8}.

These thirty$-$five  groups have the general structure \cite{Mel8}
\be
\mathcal B =  \mathcal S(A, K) \rtimes_{T} G,
\ee
where $A$ is a locally compact space, $K$ is the field of real or complex numbers,
and $G$ is the group of global, conformal, orientation$-$preserving transformations
of the space  $A$   onto itself. $\mathcal S(A, K)$  is the space of
supertranslations, i.e., of generalized position$-$dependent translations.
The supertranslations $\mathcal S(A, K)$
 are $K-$valued functions defined on $A$, and they constitute a normal abelian
 subgroup of $\mathcal B.$  Moreover, $\mathcal S(A, K)$ has a vector space structure, over the
 field $K$, under point$-$wise addition and scalar multiplication.

 In the case of $B(2,1)$ $A$ is the circle $S^{1}$, $K$ is the field of real numbers, $G$ is the group
 $ S L(2,R)$
  of all real $2\times 2$ matrices with determinant one,
  and $\mathcal S(A, K)$ is  the space  of
  real$-$valued
functions defined on $P_{1}(R)\simeq S^{1}$.
$P_{1}(R) $ is the one$-$dimensional real projective  space (the circle quotient the antipodal map).
Alternatively, as it is explained in Section \ref{sec1},
in a different realization of $ B(2,1)$,
 $S(A, K)$ is  the space  of  ``even'' real$-$valued defined on $ S^{1}$.

Ashtekhar $\&$ Hansen, in a unified treatment of asymptotic flatness in both null and spacelike directions for real lorentzian 4$-$dim space$-$times \cite{ass},
derived a BMS$-$like group, the \it Spi \normalfont group, which acts on the spacelike analogue of ``null infinity" $\Im^{+}$, the
``spacelike infinity". ``Spacelike infinity" is a 4$-$dim manifold, as opposed to ``null infinity" $\Im^{+}$ which is a 3$-$dim manifold.
One of the  thirty$-$five generalizations  of $B(2,1)$, is the group $B_{S}(2,1)= \mathcal S(S^{1} \times I, R) \rtimes_{T} SL(2,R)$,
which is the analogue of the  \it Spi \normalfont group  in
3$-$dim space$-$times. $I$ is an open interval of $R,$ and $\mathcal S(S^{1} \times I, R)$ is
the space of real$-$valued
functions defined on the product $S^{1} \times I$ which is homeomorphic to a cylinder in $R^{3}$.
In the case of 3$-$dim space$-$times ``spacelike infinity" is a 3$-$dim manifold, whereas ``null infinity" is a 2$-$dim manifold.

By following Penrose \cite{Pen1,Pen2,Pen3} we can interpret $B(2,1)$ as a transformation group $B(2,1) \times \Im^{+}$
of the ``future null infinity" of the Lorentzian 3$-$dim space$-$times involved.
Furthermore, we can  give   a geometric structure to $\Im^{+}$, the ``strong conformal geometry" for which the
transformation group is precisely the group of automorphisms. 
The assignment of the ``strong conformal geometry" to the
2$-$dim
$\Im^{+}$, in the case of 3$-$dim space$-$times,
follows Penrose's original  beautiful construction \cite{Pen1,Pen2,Pen3}
of the ``strong conformal geometry" in the case of 4$-$dim space$-$times.

However, the definition of the ``strong conformal geometry",  in the case of 3$-$dim space$-$times, does not come as an immediate corollary of Penrose's original construction
 of the ``strong conformal geometry",
in the case of 4$-$dim space$-$times, simply by
following through Penrose's construction and by
reducing on the way the number of spatial dimensions by one.
There are some subtleties involved when we pass from
4 space$-$time dimensions to 3 space$-$time dimensions.
In particular some care is needed,  in 3 space$-$time dimensions, in the appropriate definition
of the notion of the ``null angle",
introduced by Penrose \cite{Pen2,Pen3}, in order to establish
the ``strong conformal geometry".
Here we do not elaborate further on this issue but we rather
restrict our attention  to the representation theory of $B(2,1)$ and its subtle underpinnings.

As it is explained in Section \ref{sec3} in
order to find the IRS  of any BMS group $B$,  as well as the IRS of any of its analogues,
 in any number $d\geq3$
of space$-$time dimensions, we cannot
apply Wigner$-$Mackey's theory as it is.
We need to extend it.
The required extension, is reduced  to the solution of three problems, described in detail in Subsection
\ref{extension}. These three problems are named the First, the Second, and the Third problem, and they are
correspondingly, the ``little groups'' problem, the irreducibility problem, and the exhaustivity problem.

In Subsection
\ref{little}
we find a solution to the First problem. The solution
applies to any topology $ B$  and its analogues come equipped with, in any number $d\geq3$
of space$-$time dimensions. The solution to the Second problem, given in
Subsection \ref{solutions}, applies to the Hilbert
and to the nuclear topologies. Finally, the solution to the Third problem, given in
Subsection \ref{solutions}, applies to the Hilbert
 topology only.
Therefore solution to all three problems
 posed by  the
required extension of Wigner$-$Mackey's theory are
given in the Hilbert topology, and consequently, a complete list of IRS of $B(2,1)$
can be constructed in this topology.


The extension of Wigner$-$Mackey's   theory we introduce in Section \ref{sec3}
in order to find the IRS of $B(2,1)$
is not
applicable only to $B(2,1)$
but it is also
relevant and
applicable to the thirty$-$four generalizations \cite{Mel8} of $B(2,1)$, both real in all signatures and in complex space$-$times.
In fact the relevance and applicability of the
extended theory
permeates also the representation theory of the original
BMS group $B$, as well as the representation theory of its forty$-$one generalizations \cite{mac1},
both real in all signatures and in complex space$-$times.

In fact by using the solutions to the
three problems in the Hilbert topology,
and thus the extended
Wigner$-$Mackey's   theory given in Section \ref{sec3},
 we can construct, in the Hilbert topology,
 all
the IRS of any BMS group $B$,  as well as the IRS of any of its (supersymmetric \cite{shaw}) analogues,
 in any number $d\geq3$
of space$-$time dimensions.

In the other
topologies for $B$, described in
Section \ref{top}, the
solution to the
``little groups'' problem
given in Subsection \ref{little},
allows us to
pursue Wigner-Mackey's
inducing construction and obtain
a list of representations for
$B.$ We do not know if the
representations so obtained
are irreducible, and moreover, we do not know
 if  the inducing construction gives
all the representations of $B.$

In order to find if
the representations  are
irreducible we need to solve in each case
(apart from the Hilbert topology case and the nuclear topology case) the irreducibility problem which is
stated in Subsection \ref{extension} and solved in Subsection
\ref{solutions} for the Hilbert and nuclear topologies.
Furthermore, in order to find if
we obtain all the representations  of $B$
by using Wigner$-$Mackey's inducing construction
we need to solve in each case
(apart from the Hilbert topology case)
the exhaustivity problem which is
stated in Subsection \ref{extension} and solved in Subsection
\ref{solutions} for the Hilbert topology.




To the best of our knowledge,
the  group theoretical approach  to quantum gravity advocated here, is
the only approach, in which a complete list of IRS of the appropriate
symmetry group, i.e., of $B$ and of
 any of its analogues,
 in any number $d\geq3$
of space$-$time dimensions,
is constructed in a given (i.e. in the Hilbert)
topology.

The other approaches we are aware of
\cite{Isham,IshKak0,IshKak,Bar3,Bar4} are afflicted
by the inapplicability of the
Wigner$-$Mackey's inducing construction
which manifests itself in
the appearance of the ``little groups'' problem, the irreducibility problem, and the exhaustivity problem,
which persistently occur, in this or the other form,
in any group theoretical approach to
quantum gravity.

The main results of the representation theory  are:
The
IRS of
$B(2,1)$
are induced
from IRS of \it compact \normalfont ``little groups''.
The ``little groups''
are of two types:
1. Infinite connected Lie groups. 2. Non$-$connected finite discrete groups.
The  non$-$connected finite discrete ``little groups'' are
cyclic groups of even order, which are symmetry groups of  regular
polygons in ordinary euclidean 2$-$space.
The inducing construction is exhaustive
notwithstanding the fact that $B(2,1)$
is
\it not \normalfont
locally compact in the employed Hilbert topology.


A terminological remark is here in order. Whenever, hereafter, mention is made of a representation of a group, it  will
always be taken to mean  \it strongly continuous \normalfont representation. If $\mathcal G$ is a topological group we say
that a representation $\mathcal U$ of $\mathcal G$
on a Hilbert space $\mathcal H$
is strongly continuous if $g_{n} \rightarrow g$, $n=1,2,3,... $,  in $\mathcal G$, implies $\| \mathcal U(g_{n})\psi \rightarrow \mathcal U(g)\psi   \|   $   for all $\psi \in \mathcal H.$

This paper is organized as follows.
In Section 2 relation to other approaches is
discussed.
In  Section 3 $B(2,1)$ is introduced.
In Section 4 the bare essentials of  Wigner$-$Mackey's theory are given. In Section 5 an extension of Wigner$-$Mackey's theory, necessary to construct the IRS of $B(2,1),$
and in fact the IRS of any BMS group $B$,  as well as the IRS of any of its (supersymmetric) analogues,
 in any number $d\geq3$
of space$-$time dimensions,
is developed.  In Section 6 a heuristic proof of the compactness of the ``little groups'' is given, and
 it is pointed out that besides the Hilbert topology, considered in this paper, there are other
``reasonable'' choices for the topology
of the supertranslation space. In Section 7
the necessary data in order to construct the operators
of the IRS of $B(2,1)$ are given.

\section{ Relation to other approaches }
\label{other}

\noindent
The BMS group $B$
can be derived, in general \cite{Hol,Wald,Tan1,Tan2,Tan3}, in any number $d \geq 3$ of space$-$time dimensions,
by using the four methods \footnote{There is also Bramson's derivation
\cite{Bram} who defines the BMS group $B$ to be the group of
mappings of ``cone space'', introduced by him,
 onto itself which acts linearly on the space
of position vectors and which preserves the norm of each position vector.
Bramson's definition of $B$ is illuminating and interesting but is not relevant to this study and so we do not include it in the methods of derivation of $B.$ } specified below.
The order of exposition reflects the chronological order in which they appeared
\cite{Bondi,Pen1,mac8,Ger}.

\begin{enumerate}
\item{As a pseudo$-$group of local diffeomorphisms
which preserve the asymptotic form of the
Bondi$-$Van Der Berg$-$Metzner
class of
metrics \cite{Bondi}
of Lorentzian radiating space$-$times which are asymptotically
flat in lightlike future directions.}
\item{As
  an exact global transformation group $B \times \Im^{+} \longrightarrow \Im^{+}$ of the ``future null boundary" $\Im^{+}$ of the
Bondi$-$Van Der Berg$-$Metzner space$-$times. This group is defined to be
the group of automorphisms of the ``strong conformal geometry",
 a geometric structure  given by Penrose \cite{Pen1}, to the ``future null boundary" $\Im^{+}$ of the space$-$times concerned.
}

\item{As the group which results from the
Poincare group
when the translations,
which can be put into bijective correspondence with a finite parameter family of homogeneous  functions of degree one,
are replaced by an appropriate infinite parameter family of homogeneous functions of degree one \cite{mac8}.}

\item{As a group of diffeomorphisms of the ``future null boundary" $\Im^{+}$ of the
Bondi$-$Van Der Berg$-$Metzner space$-$times which leaves invariant the
``universal structure'' \cite{Ger}. The ``universal structure'' is
equivalent to the ``strong conformal geometry" introduced by Penrose.
The key point in this method is that the invariance
of the  ``universal structure'' under the
action of $B$ on $\Im^{+}$ is expressed not in terms
of the elements of the group $B$ itself, but instead,
\it
in terms
of the linearization of the group $B$ close to its
identity element, i.e., in terms of the generators
of the Lie algebra of the group $B$.
}

\end{enumerate}

The three first methods of derivation
give the group $B$, whereas the fourth method
gives the Lie algebra of $B$. A word of caution is here in order.
Both $B$ and its (supersymmetric) analogues in any number $d\geq3$ of space$-$time dimensions are infinite dimensional
Lie groups. For  infinite dimensional
Lie groups, the exponential map, when it exists, from  a
neighbourhood of 0 in the Lie algebra of $B$ to a
neighbourhood of $e$ in $B$, is \it not \normalfont surjective.
Thus exponentiation of the Lie algebra of $B$ does not give, in general,
the connected component of the identity of $B$.
Therefore any conclusions drawn for $B$ by studying its Lie algebra
should be drawn
with this word of caution in mind.

One can also derive the variants of $B$,
as well as their supersymmetric counterparts,
in any number $d\geq3$ of space$-$time dimensions,
with the use of the four methods stated above, by appropriately
modifying in each case, the class of metrics in the first
method, the notion of ``strong conformal geometry" in the second method,
the appropriate infinite parameter family of homogeneous functions of degree one in the
third method, and finally, the notion of ``universal structure"
in the fourth method.
\noindent

Each method
has its own merits and drawbacks. For example,
it appears \cite{Wald}, that
in odd
space$-$time dimensions, the second and the fourth method of derivation
are not applicable in the case of \it radiative \normalfont space$-$times.
This restriction is not applicable to the case of three space$-$time
dimensions we are studying in this paper,
because in three space$-$time
dimensions
there are no propagating degrees of freedom, i.e., there is no gravitational
radiation.




Here we introduce
the analogue $B(2,1)$
of 
the BMS group $B$
in 3 space$-$time dimensions
by using
for the first time the algebraic way of derivation.
To the best of our knowledge,
the asymptotic symmetry group in 3 space$-$time
dimensions
was firstly considered
in \cite{Bij}, a work which is entirely classical in aims and scope.
In particular  no IRS were considered in this study.



In this investigation the authors studied
gravitational waves with a space-translation Killing field. In this case, they noticed, the 4$-$dimensional Einstein vacuum equations are
equivalent to the 3$-$dimensional Einstein equations with certain matter sources.
The key point
in their study
which determined the structure of the asymptotic
symmetry group they derived,
is the following. The vacuum gravitational wave solutions
they considered in 4 space$-$time dimensions are not asymptotically flat either at spatial
or null infinity. This led to weaker fall off conditions
for the matter sources in the equivalent system of
Einstein equations with matter sources.

The weaker fall off conditions led in turn to the enlargement of the asymptotic
symmetry group they derived.
It led in particular to
the enlargement of $SL(2,R)$
to the full group $Diff(S^{1})$ of the orientation preserving diffeomorphisms of the
circle.
 $SL(2,R)$ acts on $P_{1}(R)\simeq S^{1} $, the one-dimensional real projective  space (the circle quotient the antipodal map), this
 is going to be of interest in  Subsection \ref{sec2.3}.
As a result of this, the group obtained by the authors, by using the fourth method of derivation, is not
the semidirect product of supertranslations, i.e., of the additive group of functions on  the
circle, with $SL(2,R)$, which is  the group $B(2,1)$ considered in this
paper, but it is instead the semidirect product of the supertranslations
with $Diff(S^{1})$.

The analogue of the BMS group $B$
in 3 space$-$time dimensions with the first method of derivation was given in \cite{Bar2} and \cite{Bar3}.
In \cite{Bar1} the authors 
advocated the view put forward e.g. in
\cite{Str,Str1,Graham,Aha,Boer,Sol}, namely that 
physics should be constrained by infinitesimal 
symmetry transformations that are not necessarily 
globally well defined.

With this motivation
 the authors initiated a research program
in \cite{Bar1} whose objective is to materialize this
thesis in Einstein's G.R. in asymptotically flat space$-$times.
Firstly they considered the 4$-$dim case \cite{Bar1,Bar2}, and then 
they examined the  3$-$dim case \cite{Bar2,Bar3,Bar4}.


In particular, in three space$-$time dimensions, the authors 
studied \cite{Bar3,Bar4}, by using Wigner$-$Mackey's theory,
the projective representations  of BMS$_{3}$, the semidirect product of
$Diff(S^{1})$  with the supertranlations, the additive group 
of functions on the circle. 
Thus both factors of  BMS$_{3}$ are infinite dimensional. 
There is no overlap 
between this paper and \cite{Bar3,Bar4} because the 
group $B(2,1)$ we are considering in this paper is different from the group BMS$_{3}$ whose representation theory is undertaken in \cite{Bar3,Bar4}.

The representation theory of  BMS$_{3}$ is also afflicted 
by the inapplicability of the
Wigner$-$Mackey's inducing construction
which persistently occurs, as we pointed out in Section 1, in this or the other form,
in any group theoretical approach to
quantum gravity.

In the case of BMS$_{3}$ the  
inapplicability of the
Wigner$-$Mackey's inducing construction
 manifests itself in
the appearance of the exhaustivity problem 
and in the construction of quasi$-$invariant$-$measures
in the appropriate orbits \cite{Bar4}. No solution is 
given by the authors to these problems and are left for
future  consideration \cite{Bar4}.   
In the case of $B(2,1)$ solution to the exhaustivity problem
in the Hilbert topology is given in Subsection \ref{solutions}
and quasi$-$invariant$-$measures are constructed in Subsection
\ref{invmeas}.

A last remark is here in order regarding 
$B(2,1)$ and BMS$_{3}$. $B(2,1)$ is a proper subgroup of BMS$_{3}$. 
However, this does not render the study of the representations
of $B(2,1)$ superfluous. In general, the following holds: Let $K$ be a group and $H$ be a subgroup of $K$. Let $T(k)$ be
an irreducible representation of the group $K$. Let $T_{H}(k)$ be the restriction of $T(k)$ on the subgroup
$H$. In general, the representation $T_{H}(k)$ is not irreducible. Moreover, there are irreducibles of $H$ which
cannot be extended to the whole group $K$ .

As we pointed out, nuclear topology IRS of $B$
appear to allow scattering states, i.e., unbounded
sources, possibly with infinite energy, and also
distributional metric solutions to Einstein's equations
\cite{mac6}.
It is interesting to note that a completely different
group theoretic approach to quantum gravity has been developed by Isham \cite{Isham,IshKak0,IshKak}. In this
approach IRS of the infinite dimensional canonical group
$\mathcal C$ for quantum gravity are studied. This fascinating work
also arrives, independently and from another point of
view, at distributional metrics. In Isham's work, however,
these metrics occupy the centre of the stage.

$\mathcal C$ has the semidirect product structure
\be
\mathcal C= \mathcal S \rtimes_{T} G,
\ee
where $\mathcal S$ is an Abelian subgroup of $\mathcal C,$
and $G$ is a \it not \normalfont locally compact group. The
linear representation of $G$ on $\mathcal S$ which specifies
this semidirect product is not relevant here.
Isham does not succeed in obtaining a complete
list of IRS of the canonical group. His efforts in doing so are hindered by two problems:
\begin{enumerate}
\item{The exhaustivity problem: The canonical group is not locally compact,
Mackey's theorems (see e.g. \cite{Mackey1}) ensuring the
exhaustivity of the IRS constructed by inducing cannot
be invoked, and, as a result, there is no clear idea,
of how many IRS of the canonical group can be obtained
by Wigner$-$Mackey's inducing construction.}
\item{The problem, which proves to be the central mathematical problem in Isham's approach, of
    finding $G$$-$quasi$-$invariant measures on the
    various $G$$-$orbits in  $ \mathcal S^{\prime}$,
    $ \mathcal S^{\prime}$ is the topological dual of $ \mathcal S$ . The actual construction of quasi$-$invariant measures in Isham's approach cannot
    be attained by following the usual method of
   projecting down the Haar measure from $G$ to
   the $G$$-$orbit $\mathcal O \simeq G/L_{f}$,
   $L_{f}$ is the ``little group'' of a representative
   point $f$ of $\mathcal O,$ since $G$ is not locally
   compact and hence has no Haar measure.
   }
\end{enumerate}

The exhaustivity problem in our approach is described in detail in Subsection \ref{extension}  and its solution,
in the Hilbert topology, is given in Subsection \ref{solutions}. One key difference between Isham's approach
and our approach is that in our approach the group $G$ is locally compact. As a result,
the method of
   projecting down the Haar measure from $G$ to
   the $G$$-$orbit $\mathcal O \simeq G/L_{f}$ can be invoked, and
$G$$-$quasi$-$invariant measures on the orbits $G/L_{f}$   can be constructed.
For the case of $B(2,1)$ this construction is carried out in Subsection   \ref{invmeas}.

Both problems in Isham's approach have to do with the applicability of Wigner$-$Mackey's induced
representation theory. In our approach in order to apply
Wigner$-$Mackey's theory we need to extend it. As we
stated in the previous Section the required extension
is given in Subsection \ref{extension} and it is reduced to the solution of three problems, and the solution to the three
problems is examined in Subsection \ref{solutions}.

\section{ The group $B(2,1)$ }

\label{sec1}
We turn now to the study of $B(2,1),$
the analogue of $B$ in three space$-$time dimensions.

\subsection{The group $B^{2,1}(N^{+})$}

\noindent
Recall that the $2+1$
Minkowski space is the vector space $R^{3}$ of row vectors with 3
real components, with the inner product defined as follows. \ Let
$x,y\in R^{3}$ have components $x^{\mu }$ and $y^{\mu }$
respectively, where $\mu =0,1,2$. \ Define the inner product
$x.y$ between $x$ and $y$ by
\begin{equation}
x.y=x^{0}y^{0}-x^{1}y^{1}-x^{2}y^{2}.
\end{equation}

Then the $2+1$ Minkowski space, sometimes written
$R^{2,1}$, is just $R^{3}$ with this inner product.
The ``2,1'' refers to the one plus and two minus signs in the inner product.
Let $SO(2,1)$ be the (connected component of the identity element of the)
group of linear transformations preserving the inner product. \ Matrices
$\Lambda \in SO(2,1)$ are taken as acting by matrix multiplication from the
right, $x\longmapsto x\Lambda ,$ on row vectors $x\in R^{2,1}$. \

The future null cone $ N^{+} \subset R^{2,1}$ is just the set of nonzero vectors
with zero length and $x^{0}>0$:
\begin{equation}
 N^{+}=\left\{ x\in R^{2,1}| x.x=0, x^{0}>0
\right\} .
\end{equation}
Let $R^{*}_{+}$ denote the multiplicative group of all
\it{positive}
\normalfont
real numbers. \ Obviously, if $x\in N^{+}$, then
$tx\in N^{+}$ for any $
t\in R^{* }_{+}$. \ Let $F_{1}(N^{+})$ denote the vector space (under
pointwise addition) of all functions $f:N^{+}\rightarrow R$ satisfying
the homogeneity condition
\begin{equation}
\label{hom}
f(tx)=tf(x)
\end{equation}
for all $x\in N^{+}$ and $t\in R^{*}_{+}$.

Define a representation $T$
of $SO(2,1)$ on $F_{1}(N^{+})$ by setting, for each $x\in N^{+}$ and $\Lambda \in SO(2,1)$,
\begin{equation}
(T(\Lambda )f)(x)=f(x\Lambda ).
\end{equation}
Now let $B^{2,1}(N^{+})$ be the semidirect product
\begin{equation}
\label{tvyrikolasexy}
B^{2,1}(N^{+})=F_{1}(N^{+}) \rtimes_{T}
SO(2,1).
\end{equation}
That is to say, $B^{2,1}(N^{+})$ is, as a set, just the product
$F_{1}(N^{+})\times
SO(2,1)$, and the group multiplication law for pairs is
\begin{equation}
(f_{1},\Lambda _{1})(f_{2},\Lambda _{2})=(f_{1}+T(\Lambda _{1})f_{2},\Lambda
_{1}\Lambda _{2}).
\end{equation}

Let $R^{*}_{+}$ be the multiplicative group of positive real numbers.
The orbits of the dilatation action
\be R^{*}_{+} \times N^{+} \rightarrow N^{+}; \ (t,x) \mapsto tx,\ee $t \in R^{*}_{+}, \ x \in N^{+},$
are just the open half lines in $N^{+}$ from the origin of $R^{2,1}$. The projective null cone $P( N^{+})$
is precisely the space of orbits of this action.

We can replace the set of all homogeneous functions, $F_{1}(N^{+})$,
on $N^{+}$, by the set $F(P(N^{+}))$ of all ``free'' functions on $P(N^{+})$, since the homogeneity constraint fixes the
behaviour of functions $f \in F_{1}(  N^{+})$  on these half lines. This gives a bijection
\be
\label{bi1}
F_{1}(N^{+}) \leftrightarrow  F(P(N^{+}))
\ee
by means of which we can find an expression for the representation $T$ in the space $F(P(N^{+}))$, giving us a new realization
\be
\label{grouppr}
B^{2,1}(N^{+})=F(P(N^{+})) \rtimes_{T}SO(2,1).
\ee

Each half$-$line  $t  x,$  $ t \in R^{*}_{+}, \ x \in N^{+},$
intersects the 2$-$plane $x^{0}=1$ exactly once; $x^{0}$ is the zero component of $x$. Since $x=(x^{1},x^{2},x^{0}) \in N^{+}$, $x$  is a null vector, therefore, $(x^{1})^{2}+(x^{2})^{2}=(x^{0})^{2}.$
We conclude that the homeomorphic type of the space of orbits  $P(N^{+})$ is a circle, in the model for $P(N^{+})$  constructed here,  given by
\be
(x^{1})^{2}+(x^{2})^{2}=1.
\ee

Rather than work with (\ref{grouppr}), it is more convenient to pass
to a ``spinor'' version of $B^{2,1}(N^{+})$ which double covers this group.

\subsection{The spinor version of $B^{2,1}(N^{+})$}
\label{dc}
\noindent
$
S
 L
 (2,R)$ is sometimes denoted by $G$ below.
Let ${\rm M}_{s}(2,R)$ be the set of all $2\times 2$ symmetric real matrices.
We define a right action of $G $ on ${\rm
M}_{s}(2,R)$,
\begin{equation} \label{axcsdfvbgtre}
 {\rm M}_{s}(2,R) \times G \rightarrow {\rm M}_{s}(2,R) \
\rm{with} \  \normalfont (\it{m}\normalfont,\it{g}\normalfont)
\mapsto \it{g}^{\top} \normalfont \it{m}\normalfont \it{g},\normalfont
\end{equation}
where the superscript ${\top}$ means transpose.

Clearly any element $\mu \in {\rm M}_{s}(2,R)$ can be parameterized as follows:
$$
\mu=\left[
\begin{array}{cc}
   x^{\rm o}-x^{1}    &   x^{2} \\
    x^{2}     &  x^{o}+x^{1}
\end{array}  \right]
$$
where $x^{\rm o},x^{1},x^{2} \in R$.
We now consider
the map $b:R^{2,1} \rightarrow {\rm M}_{s}(2,R)$ defined by \begin{equation}
\label{wyujhdcvxsertzawl} b(x)=\left[
\begin{array}{cc}
   x^{\rm o}-x^{1}    &  x^{2} \\
    x^{2}     &  x^{o}+x^{1}
\end{array}  \right],
\end{equation} where the $x^{\mu}$ are the components of $x\in R^{2,1}$.

This
map is a linear bijection, so the right action of $G$ on ${\rm
M}_{s}(2,R)$ induces a linear right action of $G$ on $R^{2,1}$.
The right action of $G$ preserves determinants:
\be
\rm{det}\normalfont(\it{g^{\top} m g})\normalfont=
\rm{det}\normalfont(\it{g}^{\top}) \normalfont
\rm{det}\normalfont(\it{m})\normalfont
\rm{det}\normalfont(\it{g})\normalfont=
\rm{det}\normalfont(\it{m})\normalfont.
\ee
Moreover, \begin{equation} \det(b(x))=x \cdot x. \end{equation}

Therefore the right action (\ref{axcsdfvbgtre}) of $G $ on ${\rm M}_{s}(2,R)$,
corresponds, via the linear bijection $R^{2,1} \leftrightarrow {\rm M}_{s}(2,R)$, to linear maps preserving lengths $x \cdot x.$ In fact, all linear maps $ \Lambda \in SO(2,1)$ are obtained in this way, and the action (\ref{axcsdfvbgtre}) gives a homomorphism
\begin{equation} \gamma : G
\rightarrow SO(2,1),
\end{equation}
which is onto, and has kernel $ Z_{2}=\{ Id,-Id \} $ in
$G$, $Id$ denoting the identity element of $G$.
Thus $ \gamma $ identifies $ G  $ as the
double cover of $ SO(2,1), $ \begin{equation} G = SO(2,1)_{c}. \end{equation}

The null cone $N^{+}$ becomes, under the identification
$R^{2,1} \leftrightarrow {\rm M}_{s}(2,R), $
$\widetilde{N^{+}}=b(N^{+})$ with
\begin{equation}
\widetilde{N^{+}}
=\left\{ \mu \in
{\rm M}_{s}(2,R),\;\mid x^{0} > 0,   \;
\det \mu=0\right\} .
\end{equation}
Thus $\mu \in \widetilde{N^{+}}
$ if and only
if $\mu $ has rank exactly $1$ and $x^{0} > 0$.
Let $\scri$ be the set of all non$-$zero real two$-$component row vectors $\bm{\sigma}$; $\scri=R^{2}- \{0\}$.
From the rank condition and the requirement $x^{0} > 0$, it follows that
\begin{equation}
\label{sp}
\mu \in \widetilde{N^{+}}
\quad \Leftrightarrow \quad \mu =\bm{\sigma}^{\top}\bm{\sigma}
\end{equation}
where  $\bm{\sigma} \in \scri$.
\noindent

Equation (\ref{sp}) gives a parametrization of
the null cone $
\widetilde{ N^{+}}
$ by
means  of spinors.
However, this parametrisation, though (by construction) onto, is not one$-$one. It is onto because \it{every}  \normalfont
$\mu \in \widetilde{N^{+}}$   can be parameterized in this form but it is not one$-$one because both $\bm{\sigma} $ and $-\bm{\sigma} $ give rise to the same $\mu \in \widetilde{N^{+}}.$


There is a natural   left action of $R^{*}_{+}$ on $\widetilde{ N^{+}}$
\be R^{*}_{+} \times \widetilde{ N^{+}} \rightarrow \widetilde{ N^{+}}; \ (t,\mu) \mapsto t\mu, \ee
$t \in R^{*}_{+}, \ \mu \in \widetilde{ N^{+}}.$
The aforementioned linear bijection
$$
b: N^{+} \rightarrow \widetilde{ N^{+}}  \
$$
is preserved by the left $R^{*}_{+}$ action on $N^{+}$
and on $\widetilde{ N^{+}}.$

Therefore the orbits of the $R^{*}_{+}$ action on $N^{+}$
are in bijective correspondence with the orbits of the $R^{*}_{+}$ action on $\widetilde{ N^{+}}.$ Consequently, with this identification, we can use as a model for the
orbits of the $R^{*}_{+}$ action on $N^{+}$, the
orbits of the $R^{*}_{+}$ action on $\widetilde{ N^{+}}$.
Let $\mathcal O=t \mu,$ $t \in R^{*}_{+},\ \mu \in \widetilde{N^{+}},$
be an orbit of the $R^{*}_{+}$ action on $\widetilde{N^{+}}.$
Let $\mathcal O_{r}$ be any representative of this orbit.
$\mathcal O_{r}$ can be parameterized in terms of
spinors, and, as (\ref{sp}) indicates,
if $\bm{\sigma} $  corresponds to $\mathcal O_{r}$ then $ -\bm{\sigma} $ also corresponds to $\mathcal O_{r}.$

We conclude that the projective space $P(\widetilde{ N^{+}})$ of the orbits of the $R^{*}_{+}$ action on $\widetilde{N^{+}}$ is
\be
P(\widetilde{ N^{+}})=P_{1}(R)= S^{1}/Z_{2} \simeq S^{1},
\ee
The fact that both $\bm{\sigma}$ and $-\bm{\sigma}$ correspond to
the \it same \normalfont orbit $\mathcal O_{r}$ provides the starting point for a more rigorous proof \cite{Melas}
of the homeomorphic type of  $P(\widetilde{ N^{+}}).$

Consequently the spinor version $B^{2,1}(N^{+})_{c}$
of $B^{2,1}(N^{+})$ which double covers $B^{2,1}(N^{+})$ is
\be
B^{2,1}(N^{+})_{c}=F(P(\widetilde{ N^{+}}) \rtimes_{T}G,
\ee
where $G=SL(2,R).$ Interestingly enough
$F(P(\widetilde{ N^{+}})$ can be realized in two different ways
which give rise to two different isomorphic realizations of
$B^{2,1}(N^{+})_{c}.$

\vspace{0.2cm}
\noindent
$\bf 1^{st} \ realization.$ \normalfont \hspace{0.2cm}
$F(P(\widetilde{ N^{+}})$ is the space  of ``free'' real$-$valued
functions $F(P_{1}(R))$ defined on $P_{1}(R)$.
Thus in this case $B^{2,1}(N^{+})_{c}$ is realized as
\be
B^{2,1}(N^{+})_{c}=F(P_{1}(R)) \rtimes_{T}G.
\ee

\vspace{0.2cm}
\noindent
$\bf 2^{nd} \ realization.$ \normalfont \hspace{0.2cm}
Let $S^{1}$ be the curve $S^{1} \subset I$ defined by
\be
S^{1}=\{ \bm{m}  \in I | \ |\bm{m}|=1\}.
\ee
$F(P(\widetilde{ N^{+}})$ is the space  of ``even'' real$-$valued
$\beta:S^{1} \rightarrow R.$ That is, these functions satisfy the
even$-$ness condition
\be
\beta(\bm{m})=\beta(-\bm{m}).
\ee
We denote this space by $F_{e}(S^{1}).$
Thus in this case $B^{2,1}(N^{+})_{c}$ is realized as
\be
B_{e}^{2,1}(N^{+})_{c}=F_{e}(S^{1}) \rtimes_{T}G.
\ee

\vspace{0.2cm}

We can use either of these realizations to develop the representation theory of $B^{2,1}(N^{+})$.
The representation theory which has as point of departure the
first realization cannot be mapped onto the representation theory  which has as point of departure  the second realization. Each representation theory presents its own difficulties, problems and  subtleties. However, as expected, the results of the representation theory which stems from the first realization can be naturally mapped onto the results of the representation theory which stems
from the second realization. Here we restrict attention to the representation theory which originates from the first realization.



\subsection{The group $B(2,1)$}
\label{sec2.3}

\noindent
In the $1^{\rm{st}}$ realization $F(P(\widetilde{ N^{+}})$ is realized as
the space  $F(P_{1}(R))$ of ``free'' real$-$valued
functions  defined on $P_{1}(R)$.
As  a result we have the following Theorem:

\begin{theorem}
The group $B^{2,1}(N^{+})_{c}$ can be realized as
\be
\label{defb}
B^{2,1}(N^{+})_{c}=F(P_{1}(R)) \rtimes_{T} G,
\ee
with semidirect product specified by
\be
\label{actb}
(T(g)\alpha)(\sigma)=\kappa_{g}(\sigma)\alpha(\sigma g).
\ee
\end{theorem}

\vspace{0.2cm}

\noindent
$G$ is the group SL(2,$R$), $g \in G$,
$\alpha \in F(P_{1}(R))$.
Moreover, if
\begin{equation}
g=\left[ \begin{array}{lr}
   a    & b \\
   c    & d
\end{array}  \right]
,
\end{equation} then the components $ \sigma_{1},\sigma_{2} $ of $\bm{\sigma} \in R^{2}-\{0 \}$ transform
linearly under $g$, so that the ratio $ \sigma= {\sigma_{1}}/{\sigma_{2}}, \ \sigma_{2} \neq 0 $, \  transforms
fraction linearly under $g$. Writing $ \sigma
g $ \ for the transformed ratio, \begin{equation}
\label{aliki} \sigma g=
\frac {(\bm{\sigma}g)_{1}}{(\bm{\sigma}g)_{2}}=\frac
{\sigma_{1}a + \sigma_{2}c} {\sigma_{1}b+\sigma_{2}d}=
\frac { \sigma a+c}{\sigma   b+d}. \end{equation}
The ratio $ \sigma= {\sigma_{1}}/{\sigma_{2}}, \ \sigma_{2} \neq 0 $, is a local inhomogeneous coordinate of $P_{1}(R)$.
The factor $\kappa_{g}(\sigma)$ on the right hand side of (\ref{defb})
turns out \cite{Melas} to be
\begin{equation}
\label{donkey}
\kappa_{g}(\sigma)=
\frac {(\sigma b+d)^{2}+(\sigma a+c)^{2}}{1+\sigma^{2}}.
\end{equation}

In analogy to $B$, it is natural to choose a measure $\lambda$
on $P_{1}(R)$ which is invariant under
the maximal compact subgroup $ S O (2)$ of $ G $; so we choose $\lambda$ to be the standard
normalized Lesbegue measure $ {\rm d} \lambda = \frac{{\rm d} \theta}{2 \pi}$.
For physical applications, it is necessary to give $F(P_{1}(R))$
additional structure. For reasons discussed in detail in McCarthy \cite{mac1},
the supertranslation space $F(P_{1}(R))$  is restricted
to be the separable Hilbert space $L^{2}(P_{1}(R),\lambda ,R)$
of real$-$valued functions defined on
$P_{1}(R) \simeq S^{1}$; functions
square integrable
with respect to $\lambda$.
$B^{2,1}(N^{+})_{c}$ can now be realized in terms of $L^{2}(P_{1}(R),\lambda ,R)$ \cite{Melas} as follows:

\begin{theorem}
\label{omorpho} The group $B^{2,1}(N^{+})_{c}
$   can be realised as \begin{equation}
B(2,1)
= L^{2}(P_{1}(R),\lambda ,R)
\rtimes_{T} G \end{equation} with semidirect
product specified, as before, by
$$
\label{an}
(T(g)\alpha)(\sigma)=\kappa_{g}(\sigma)\alpha(\sigma g).
$$
\end{theorem}
\vspace{0.2cm}

\noindent
In the last relation
$\alpha \in L^{2}(P_{1}(R),\lambda ,R),$ and $\kappa_{g}(\sigma)$
is given, as before,
by (\ref{donkey}).
We denote our final realization of our group
by $B(2,1)$ to
distinguish it from the previous realizations $B^{2,1}(N^{+})$ and $B^{2,1}(N^{+})_{c}$.

Reisz$-$Fr$\acute{e}$chet theorem for Hilbert spaces \cite{Rudin}
states   that the topological dual of a Hilbert space can be identified
with the Hilbert space itself, so that we
have $ {L^{2}}^{'}(P_{1}(R),\lambda ,R) \simeq
L^{2}(P_{1}(R),\lambda ,R).$
In particular, given a continuous linear functional
 $f \in {L^{2}}^{'}(P_{1}(R),\lambda ,  R)$, we can write,
 for $\alpha \in L^{2}(P_{1}(R),\lambda ,R)$
\begin{equation}
(f,\alpha)=<f,\alpha>,
\end{equation}
where the function $f \in L^{2}(P_{1}(R),\lambda ,R)$ on the right is uniquely determined
by (and denoted by the same symbol as) the linear functional $f \in {L^{2}}^{'}(P_{1}(R),\lambda ,R)
$ on the left.

The representation theory of
$B(2,1)$
is governed by the dual
action $T'$ of $ G$ on the topological dual
$ {L^{2}}^{'}(P_{1}(R),\lambda ,R)$ of $L^{2}(P_{1}(R),\lambda ,R).$
The dual action $T'$ is defined by:
\begin{equation} <T'(g)f,\alpha>=<f,T(g^{-1})\alpha>\cdot
\end{equation}
A
short calculation
gives
\begin{equation} \label{tzatziki}
(T'(g)f)(\sigma)=\kappa_{g}^{-2}(\sigma)f(\sigma g).
\end{equation}
Now, this action $T'$ of $ G$
on ${L^{2}}^{'}(P_{1}(R),\lambda ,R)$,
given explicitly above is, like the action $T$ of
$ G$ on $L^{2}(P_{1}(R),\lambda ,R)$, continuous. The ``little group'' $L_{f}$
of any  $f \in {L^{2}}^{'}(P_{1}(R),\lambda ,R)$ is the stabilizer \begin{equation}
L_{f}=\{g \in  G
\ | \ T'(g)f=f\}. \end{equation}


\section{Wigner$-$Mackey's theory}
\label{sec2}

Our construction of the IRs of $B(2,1)$ is based on G.W.Mackey's pioneering work on group representations (see, for example, \cite{Mackey0,Mackey00,Mackey,Mackey1});
in particular, is based on an extension to the relevant infinite dimensional case
\cite{mac6,Mel5}
of his semidirect product theory.
One of the sources of Mackey's  work
was Wigner's classic paper \cite{Wigner} in which he broke new grounds in mathematics by giving the first explicit treatment of of infinite dimensional
representations of a Lie group.

We will give the bare essentials of
Wigner$-$Mackey's theory  in order to construct explicitly
the operators of the induced representations of $B(2,1).
$ We will not follow the functional analytic treatment of induced
representations but instead we shall adopt the more geometrical
fibre bundle dominated scheme. Standard references are
(\cite{Wigner},\cite{Mackey},\cite{Mackey1},\cite{Simms},
\cite{Isham} and references therein).

Let $ A $ and $ \mathcal G $ be topological groups, and let $ T $
be a given homomorphism from $ \mathcal G $ into the group of
automorphisms ${\rm Aut} (A)$ of $A.$ Suppose $A$
is abelian and
$\mathcal B = A  \rtimes_{T} \mathcal G $
is the semidirect product of $A$ and $ \mathcal G,$
specified by the continuous action $T: \mathcal G
\longrightarrow  {\rm Aut} (A).$ In the product topology of
$A \times \mathcal G,$ $\mathcal B $ then becomes a topological
group. It is assumed that it becomes a separable locally compact topological
group.

The irreducible unitary continuous representations of $A$ (characters) 
are
\noindent
continuous homomorphisms
\be
\phi:  A \rightarrow U(1)
\ee
of $ A$ into the multiplicative group of complex numbers of unit modulus. A composition law
\be
\label{com}
(\phi_{1} \phi_{2}) (\alpha) = \phi_{1}(\alpha) \phi_{2}(\alpha)
\ee
on the characters and a multiplication by scalars
\be
(\kappa \phi)(\alpha)=\kappa \phi(\alpha), \ \kappa \in R,
\ee
turns the characters into a vector space over the reals.
In particular the composition law (\ref{com}) turns the characters
into an abelian group, the dual group $\hat{A}$ of $A$.

Writing the character $\phi$ as
\be
\label{chi}
\phi(\alpha)=e^{i f(\alpha)},
\ee
one easily finds \cite{Melas}
that the real$-$valued function $f$
is a continuous linear functional on $ A$, that is, $f$ belongs to the topological dual $ A^{\prime}$ of $ A$. The map
\begin{eqnarray}
j:  A^{\prime} & \rightarrow & \hat{ A}, \ \rm{given \ by} \\
\label{isoj}
j(f) & = & e^{i f}
\end{eqnarray}
provides an isomorphism between $\hat{ A}$ and $ A^{\prime}$, as real vector spaces, and in particular as abelian groups.
We shall identify, if required, the sets
$\hat{ A}$ and $ A^{\prime}$.

The action $T$ of $\mathcal G $ on $ A $
induces a dual action on $ \hat{A}.$
Indeed, any bijective map $b: A \longrightarrow A$,
induces a map $\hat{A} \longrightarrow  \hat{A}$,
$\phi  \longmapsto b \phi$, defined by
\be
\label{indact}
(b \phi)(a)=\phi(b^{-1}a).
\ee
It is to be noted that the bijective correspondence
\be
\label{bc}
\hat{ A} \leftrightarrow  A^{\prime}
\ee
is preserved pointwise by the $\mathcal G$ action
if and only if the action of $\mathcal G$
on $ A^{\prime}$ is defined by
\be
\label{da}
(gf)(\alpha)=f(g^{-1}\alpha).
\ee
In this way an action $\alpha \longmapsto  h   \alpha h^{-1}   $      of $\mathcal B$ on $A$ induces an action
$\phi \longmapsto h \phi$ of $\mathcal B$ on $\hat{A}$,
\be
\label{actionaction}
(h \phi)(\alpha)= \phi (h^{-1} \alpha h).
\ee
\noindent

For a  given character
$ \phi \in \hat{A},$ the largest subgroup $L_{\phi} $
of $\mathcal G $ which leaves $\phi $ fixed is called
the ''little group'' of $ \phi $, and the set of characters which can be
reached from $ \phi $ by the
$  \mathcal B $ action is called the orbit
of $ \phi $, denoted $ \mathcal B \phi.$
The largest subgroup of $ \mathcal B $ which leaves $\phi $
fixed is the semidirect product
$ \mathcal B_{\phi}=
A \rtimes_{T} L_{\phi }  \;$.
$\mathcal B_{\phi} $ is a closed subgroup of
$ \mathcal B. $

Let U be a continuous irreducible unitary representation
of the ''little group''
$ L_{\phi} $ on a Hilbert space $ D. $
Then
\be
\phi U:  (a,l) \longmapsto \phi(a) U (l),
\ee
where $a \in A $ and $ l \in L_{\phi}, $
is a continuous unitary irreducible representation of the group
$ \mathcal B_{\phi}=
A \rtimes_{T}
L_{\phi } $
in $ D.$

Let $ \mathcal B_{\phi} \rightarrow \mathcal B \rightarrow
\mathcal B / \mathcal B_{\phi} $ be the principal fibre bundle
with fibre $\mathcal B_{\phi},$ base space the coset space
$ \mathcal B / \mathcal B_{\phi} $ and the smooth projection
map $ \pi : \mathcal B \rightarrow \mathcal B / \mathcal B_{\phi} $ is the usual function $ \pi(h):= h \mathcal B_{\phi} $ which associates with each $ h \in  \mathcal B  $
the coset $ h \mathcal B_{\phi} $ to which it belongs. $\;
\mathcal B_{\phi}$  acts upon $ D $  (on the left) via
\begin{eqnarray}
\phi U &  :  & \mathcal B_{\phi}  \longrightarrow {\rm Aut}(D), \nonumber \\
&& (a,l) \longmapsto \phi(a) U(l).
\end{eqnarray}

There is a right action of $ \mathcal B_{\phi} $ on $\mathcal B
\times D $ defined by
\begin{equation}
\label{kguhv,hliytgl7ytgyukltgol7g}
(h,l)h_{\phi} \quad := \quad (hh_{\phi},h^{-1}_{\phi}v), \quad h_{\phi} \in
\mathcal B_{\phi}.
\end{equation}
The bundle with fibre $D, $
associated to the principal bundle
$ \mathcal B_{\phi} \rightarrow \mathcal B \rightarrow
\mathcal B / \mathcal B_{\phi},$
has
total space
$ \mathcal B \times_{{\mathcal B}_{\phi}}  D,$ 
which is defined to be the orbit space of $ \mathcal B \times D $
under the action (\ref{kguhv,hliytgl7ytgyukltgol7g}), and
projection map
\begin{eqnarray}
\pi_{D}  &  :  & \mathcal B \times_{\mathcal B_{\phi}} D
\longrightarrow \mathcal B/ \mathcal B_{\phi}, \nonumber \\
&& [h,v] \longmapsto \pi(h),
\end{eqnarray}
where $ [h,v] $
denotes the equivalence class of $ (h,v) \in \mathcal B \times D $
under (\ref{kguhv,hliytgl7ytgyukltgol7g}).

That $ \pi^{-1}_{D} (h \mathcal B _{\phi}) $
really is a copy $D_{h{\mathcal B}_{\phi}} $ of $ D $
can be seen by noting the existence of the map
\begin{eqnarray}
i  &  :  & D  \longrightarrow D_{\pi(h)}, \nonumber \\
&& v \longmapsto [h,v],
\end{eqnarray}
which is a diffeomorphism for each $ h \in \mathcal B. $

Therefore, for {\it each} (continuous unitary irreducible)
representation $U$ of the ''little group'' $L_{\phi}$
and for {\it each} character $ \phi $ we
have a bundle
$ D \rightarrow \mathcal B \times_{{\mathcal B}_{\phi}}  D
\rightarrow  \mathcal B / \mathcal B_{\phi}$
with fibre $ D,$
total space
$ \mathcal B \times_{{\mathcal B}_{\phi}}  D,$
and base space
$\mathcal B / \mathcal B_{\phi}.$
The bundle
$ D \rightarrow \mathcal B \times_{{\mathcal B}_{\phi}}  D
\rightarrow  \mathcal B / \mathcal B_{\phi}$
depends, up to $ \mathcal B $$-$isomorphism, only on the orbit
$\mathcal B \phi
$
and not on the choice
of $ \phi. $ There is a natural bijection
\be
\mathcal B / \mathcal B_{\phi} \longrightarrow \mathcal B \phi,
\ee
$h \mathcal B_{\phi} \longrightarrow h \phi$ which will even be a homeomorphism \cite{Bour} (Appendix I), under quite general conditions.
 We
 will
 often identify $\mathcal B / \mathcal B_{\phi}$ and $\mathcal B \phi$.

Since $A$  acts trivially on $\hat{A}$
the orbit $\mathcal B \phi$ is identical to the
orbit $\mathcal G \phi$.
Moreover,  the map
\be
\mathcal G/L_{\phi} \longrightarrow \mathcal B \phi, \quad g  L_{\phi} \longmapsto h \phi,
\ee
where $g \in \mathcal G,$ and $h=(\alpha, g), \ \alpha \in A,$ is
also a bijection.
For this reason, we shall also identify the sets $\mathcal G/L_{\phi}$ and $\mathcal B \phi$. To conclude, the following bijections hold
\be
\label{omg}
\mathcal B / H_{\phi} \longleftrightarrow \mathcal B \phi
 \longleftrightarrow \mathcal G/L_{\phi} \longleftrightarrow \mathcal G \phi,
\ee
and, for this reason, we shall identify any two of
the sets $\mathcal B / H_{\phi}, \ \mathcal B \phi, \ \mathcal G/L_{\phi}$, and $\mathcal G \phi$.

It is well known (see e.g. \cite{Simms}, p.45) that
when $\mathcal G$ is locally compact and $L_{\phi}$
is a closed subgroup of $\mathcal G$, then there is
a unique non$-$zero invariant measure class M on
$\mathcal G/L_{\phi}$. For each measure $\mu$
in M and each $g \in \mathcal G$ the measure $\mu_{g}$:
\be
\mu_{g}(E)=\mu(g^{-1}E),
\ee
where $E$ is a Borel subset of $\mathcal G/L_{\phi}$, is also in $M$.

For each $ \mu \in M $ let
\begin{eqnarray}
H_{\mu} &  = &  \left  \{
\psi \mid \psi
\quad {\rm a} \quad {\rm Borel} \quad {\rm section}
\quad {\rm of} \quad
D \rightarrow \mathcal B \times_{{\mathcal B}_{\phi}}  D
\rightarrow  \mathcal B / \mathcal B_{\phi}, \quad {\rm and}
\right . \nonumber \\
&& \left .
\quad
\int_{{\mathcal G}/L_{\phi}}<\psi(p),\psi(p)>{\rm d} \mu (p) < \infty
\right  \},
\end{eqnarray}
where $ p \in {\mathcal G}/L_{\phi} $ and $ <\psi(p),\psi(p)>$
denotes the inner product in the Hilbert space $\pi_{D}^{-1}(p)=D_{p}$
which is diffeomorphic to $ D $.
With the inner product
\be
< \psi , \phi > = \int_{ \mathcal G / L_{\phi}}
< \psi(p) , \phi(p) > {\rm d} \mu(p),
\ee
and with sections identified if they differ only on a set of $ \mu$$-$measure
zero, $H_{\mu}$ is a Hilbert space.

Define an action of $\; \mathcal B \;$ on $\; H_{\mu} \;$ by
\be
\label{ukygvkhvkyutfkyufvmhgv}
(h \psi )(p)=
\sqrt{ \frac {{\rm d} \mu_{h}}{{\rm d} \mu}(p)} h( \psi (h^{-1}p)),
\ee
where $ h \in \mathcal B,  p \in \mathcal B / \mathcal B_{\phi},$ and $\; \mu_{h}(E)= \mu(h^{-1} E) \;$ for any Borel subset $\; E \;$
of $\; \mathcal B / \mathcal B_{\phi}. \;$ This action is a unitary action.
The representation of $ \mathcal B $ on  $ H_{\mu} $
given by (\ref{ukygvkhvkyutfkyufvmhgv}) is called the representation of
$ \mathcal B $ induced by the representation $ U $ of $ L_{\phi}$. If $ U $ and $ W $ are equivalent representations of
$ L_{\phi} $ then they yield equivalent induced representations of
$ \mathcal B $.

If it is further assumed \cite{Mackey1}
that $\mathcal B$
is a \it{regular} \normalfont semidirect product then
\it{all} \normalfont IRS of $ \mathcal B $
are obtained by inducing from the IRS of
the ``little groups''
in
the way described above.
Regular
means that the $ \mathcal B-$orbit$-$space $ \hat A $
is nice in a measure$-$theoretic
sense: The $ \mathcal B-$orbits can be enumerated in some way,
namely, there is a Borel set in $ \hat A $ that meets each
$ \mathcal B-$orbit
exactly once.

\vspace{0.1cm}

The central results of induced representation theory are the following:
\begin{enumerate}
\item
{Given the topological restrictions on
$\mathcal B=
A \rtimes_{T} \mathcal G $
(separability and local compactness), any representation of $ \mathcal B,$
constructed by the method above,  is irreducible iff the representation
$ U $ of $ L_{\phi} $ on $ D$ is irreducible.
An
irreducible representation of $\mathcal B $ is obtained for {\it each}
$ \phi \in \hat A $ and {\it each} irreducible representation $ U $
of $L_{\phi}$.}
\item
{If $ \mathcal B=
A \rtimes_{T} \mathcal G $
is a regular semidirect product
then {\it all} of its IRS
can be obtained in this way.}
\end{enumerate}

It follows from the previous discussion that in order to give
the operators of the IRS of $B(2,1)$  explicitly
it is
necessary to give the following information
\begin{enumerate}
\item{
An irreducible unitary representation $U$ of $L_{\phi}$ on a
Hilbert space $D$ for each $L_{\phi}.$}
\item{
A $ \mathcal G$$-$quasi$-$invariant measure $ \mu $
on each orbit $ \mathcal G \phi
\approx \mathcal G / L_{\phi};$
where $L_{\phi} $ denotes the ''little group'' of the base point
$ \phi \in \hat A $ of the orbit
$ \mathcal G \phi.$
}
\end{enumerate}
To find the IRS
of
$ B(2,1)
= L^{2}(P_{1}(R),\lambda ,R)
\rtimes_{T} G$
then, it is enough to provide the information cited in 1 and 2
for each of the orbit types.

\section{Extension of Wigner$-$Mackey's theory
}
\label{sec3}

In 1939 Wigner \cite{Wigner}
gave the first explicit treatment of infinite dimensional representations of the Poincare group $P$, a \it ten$-$dimensional \normalfont Lie group. He constructed the IRS of $P$ by inducing
from the IRS of some of its subgroups, the ``little groups''. Mackey proved \cite{Mackey0,Mackey00}
that Wigner's inducing construction can be applied to any separable locally compact semidirect product
$\mathcal B = A  \rtimes_{T} \mathcal G $ specified by the continuous action $T: \mathcal G
\longrightarrow  {\rm Aut} (A).$
Moreover Mackey proved \cite{Mackey0,Mackey00} that
when the semidirect product is regular,
then \it all \normalfont
the IRS of $\mathcal B$ can be obtained by this inducing
construction.

It is noteworthy that both Wigner's work and Mackey's 
work are grounded on a classical theorem of Frobenius
\cite{Frob}
according to which every ``imprimitive'' representation of
a finite group is induced in a certain canonical fashion
by a representation of one of its subgroups. The notion of
``imprimitive'' is not relevant here.

\subsection{Need for an extension}
\label{extension}
What is relevant here is that
Mackey initiates his study \cite{Mackey} by assuming that
$A$ is abelian, and that both
 the
abelian group $A$ and the group $\mathcal G$ in the semidirect product
$\mathcal B = A  \rtimes_{T} \mathcal G $ are locally compact.
In the applications to Physics we are considering here,
it is assumed that $A$ has additional
structure, it is also a vector space,  with
vector addition being the group multiplication.

It is well known that every Hausdorff topological vector space is locally compact if and only if is finite. This implies in particular
that a Hausdorff infinite dimensional topological vector space is never locally compact. Hausdorff is a desirable mathematical property which it is  highly praised by Physicists since it allows to discern the points of the topological space under consideration in an way ideal for them. The same happens here; it is
expected and desired both $A$ and its topological dual $ A^{\prime}$ to be  Hausdorff topological spaces.

In all the  groups $
\mathcal B = A  \rtimes_{T} \mathcal G
$ we consider in the present paper,
the abelian group $A$ is also a
finite dimensional or
an infinite dimensional topological vector space,
which, in both cases, is Hausdorff in the whole class of topologies they come equipped with.
In particular, on every Hausdorff finite dimensional
topological vector space $A$ there is a unique
topological vector space structure, whereas,
in the case of
infinite dimensional topological vector spaces $A$
 there is a wide range of allowed topologies,
  stated in Section
 \ref{top}. In all these topologies $A$ is
 Hausdorff.
We conclude, that for the groups
$
\mathcal B = A  \rtimes_{T} \mathcal G
$ we consider in the present paper,
Wigner$-$Mackey's theory applies, as it is,
when $A$ is a finite dimensional vector space.

The key new feature \cite{mac1} of the original BMS group $B$,
as well as of all possible analogues of $B$, both real and
in any signature, or complex, with all possible notions
of asymptotic flatness ``near infinity'', is the enlargement
of the finite dimensional vector space of translations
to the infinite dimensional vector space of supertranslations. This turns $B$, as well as of all possible analogues of $B$,
into infinite dimensional Lie groups.

It is precisely this new feature which calls for an extension of
Wigner$-$Mac$-$\newline key's theory. The required extension needs to
 give answers to the following problems, emanating
precisely from the infinite dimensions of the supertranslation
space, which are not addressed by Wigner$-$Mackey's theory.
\begin{enumerate}
\item{\it{First problem. The
``little groups'' problem:
Calculation of the ``little groups'' when the abelian subgroup $A$ of $\mathcal B = A \rtimes_{T} \mathcal G $   is also an  infinite dimensional vector space.} \normalfont \newline
In a nutshell Wigner$-$Mackey's inducing construction  allows to build up the
representations of the whole group out of
representations of some of its subgroups, the
``little groups.''
Wigner$-$Mackey's theory applies to
semidirect products $\mathcal B,$
such that, when $A$ is also a vector space,
it is a finite dimensional vector space.
When $A$ is
  a finite dimensional vector space then
 its topological dual $ A^{\prime}$ is also a finite
 dimensional vector space.

 \hspace{0.4cm} There is a natural
 bijection
\be
\label{bij}
\mathcal G / \mathcal G_{f} \longrightarrow \mathcal G f,
\ee
$g \mathcal G_{f} \longrightarrow g f,$  $f \in A^{\prime},$
between the coset space $\mathcal G / \mathcal G_{f}$ and the orbit $\mathcal G f$ of $\mathcal G$ in $ A^{\prime}.$ $G_{f}$ is the ``little group'' of $\mathcal G $ at the point $f.$ A complete classification of the $\mathcal G$ orbits
on $A^{\prime},$
when $A^{\prime}$  is a  finite dimensional
vector space, is a fairly easy task (see e.g. \cite{Barut}, p. 516). This led
Wigner and Mackey
\cite{Wigner,Mackey0,Mackey} to
search for the ``little groups''  and
find them by classifying the $\mathcal G$ orbits
in $A^{\prime}.$ The situation changes drastically
when $A^{\prime}$ is an infinite dimensional vector space. Bijection (\ref{bij}) still holds,
 but a complete classification of the
$\mathcal G$ orbits in $A^{\prime}$ is not generally feasible. Consequently  the ``little groups'' cannot now be found by
classification of the
$\mathcal G$ orbits in $A^{\prime}$
and a different approach is required.
}

\vspace{0.2cm}
\item{\it{Second problem.
The irreducibility problem:
When $\mathcal B = A \rtimes_{T} \mathcal G $ is not locally compact it is not guaranteed
that
the representations of $\mathcal B$ obtained by Wigner$-$Mackey's inducing construction are irreducible.}
\normalfont \newline
When $A$ is an infinite dimensional vector space
$\mathcal B$ is not locally compact, at least in the class of topologies $A,$ and subsequently
$\mathcal B,$ come equipped with
in the present paper; these topologies are elaborated in Section \ref{top}.
In Section \ref{top} we concentrate on the ``reasonable'' topologies for $B(2,1)$. However similar
topological structures are eligible
for \it any \normalfont BMS group $\mathcal B$, and for \it any \normalfont of its analogues, in any number
$d$, $d\geq 3$, of space$-$time dimensions.
When $\mathcal B$ is not locally compact Mackey's theorems 14.1
and 14.2 in \cite{Mackey0},
which prove the irreducibility of the representations of $\mathcal B$ derived by the inducing construction,
do not no longer apply.

\vspace{0.2cm}
\item{\it{Third problem.
The exhaustivity problem:
When the abelian group $A$
 is also an  infinite dimensional vector space, special care is needed in order to prove that the inducing construction is exhaustive.
} \normalfont
\normalfont \newline
When $\mathcal B$ \normalfont
is a regular  semidirect product
Wigner$-$Mackey's inducing construction
is exhaustive, i.e., \it all \normalfont
IRS of $\mathcal B$ are induced
from
IRS of
``little groups''.
Regular semidirect product
$\mathcal B$ \normalfont means
that the action of $ \mathcal G$ on
$A^{\prime} $ is not pathological, and as a
result of this,
the $ \mathcal G-$orbit$-$space $ A^{\prime} $
is nice in a measure$-$theoretic
sense.

\hspace{0.4cm} When the space of $ \mathcal G-$orbits
in $ A^{\prime} $ is not nice, while
a given $ \mathcal G-$orbit and an irreducible
representation of the ''little group'' of a point
on that orbit still define an irreducible unitary
representation of $ \mathcal G,$ there will be still others not obtained by the inducing construction.

\hspace{0.4cm} A sufficient condition which
ensures the regularity of $\mathcal B$ (\cite{Mackey}, p.43, p.141),
is the existence of a
a Borel cross section for the
orbits in the action of $ \mathcal G$ on
$ A^{\prime},$ i.e., the existence of
a Borel set in $ A^{\prime} $ that meets each
$ \mathcal G-$orbit
exactly once.

\hspace{0.4cm}The real purpose of this condition
about the existence of a Borel cross section is that it implies that
$ A^{\prime} $ can have no $ \mathcal G-$quasi$-$
invariant measures $\mu$  such that the action of
$ \mathcal G$ is strictly ergodic with respect to
$\mu.$ Whenever such measures $\mu$ do exist, an
irreducible representation of $\mathcal B$ may
be associated with each that is not equivalent to
any of the IRS of $\mathcal B$ induced from the IRS of
the ``little groups''. The nature of these representations is such that one can well despair
of ever obtaining a classification for them as
complete and satisfying as that available for
regular
semidirect products.

\hspace{0.4cm}When the abelian group $A$
 is also a finite dimensional vector space,
then searching for a Borel cross section which
intersects each $ \mathcal G-$orbit in
$ A^{\prime} $ exactly once, is a fairly easy task
(see e.g. \cite{Simms}, p. 59).
However, when the abelian group $A$
 is also an  infinite dimensional vector space,
 as it is actually the case for both the original
 BMS group $B$ \cite{mac3} and its generalizations \cite{mac1}, and
the group $B(2,1)$ introduced here and its generalizations \cite{Mel8},
the
 situation is different.
 Then  $ A^{\prime} $ is also  an  infinite dimensional vector space.
We note that, as we clarified in Subsection \ref{extension}, at least in the class of topologies
elaborated in Section
 \ref{top}, when the abelian group $A$
 is also an  infinite dimensional vector space,
then both the BMS group $B$ in question and its
analogues, in any number $d$ of space$-$time dimensions,
$d \geq 3,$
is not locally compact and vice versa.


\hspace{0.4cm} As a result, in this case, it is not generally feasible to classify completely \cite{mac4,Melas} the orbits
of the $\mathcal G$ action on $A^{\prime}$, let alone
to search for a a Borel set in $ A^{\prime} $ which meets each $ \mathcal G-$orbit
 in $ A^{\prime} $ exactly once. We conclude
 that when $A$
 is also an  infinite dimensional vector space,
 then the existence of a Borel cross section
 which meets each $ \mathcal G-$orbit in
  $ A^{\prime} $ exactly once, cannot be used
  as a sufficient condition, which can be
  practically implemented, for the regularity
  of the semidirect product $\mathcal B = A \rtimes_{T} \mathcal G, $ and new, different
 sufficient conditions which ensure
 the regularity
  of the semidirect product $\mathcal B = A \rtimes_{T} \mathcal G$  need to be introduced.

\hspace{0.4cm} One more word of caution is in order
here, word of caution brought to the fore again by the infinite dimensions of $A$, when $A$ is also an  infinite dimensional vector space.
The new sufficient conditions
which need to
be introduced in order to ensure the
regularity of $\mathcal B,$
involve the
topologies of $A$ and of $A^{\prime}$.

\hspace{0.4cm}In a nutshell there is an interplay between the class of functions
in $A$, the class of functions in  $A^{\prime},$ and the topologies of
$A$ and $A^{\prime}:$ It is precisely the arbitrariness in the class of functions allowed in $ A,$ expounded in Section \ref{top},
 which permits a wide
range of choices of ``reasonable'' topologies for $ A$. The key point is that the type of functions in $ A$
determines the type of functions in
$A^{\prime}:$ Roughly speaking, the
``smoother''
the functions in $ A$ are, the ``rougher'' the generalized functions in $A^{\prime}$ turn out to be.  The functions in $A^{\prime}$ allow in turn
a variety
of choices of ``reasonable'' topologies for $A^{\prime}$. To conclude:

\begin{enumerate}
\item{When the abelian group $A$ in
the semidirect product $\mathcal B = A \rtimes_{T} \mathcal G$ is also an infinite dimensional vector space new sufficient conditions are needed to insure the regularity of $\mathcal B.$
}
\item{The new sufficient conditions entail the topologies of $A$ and of its topological dual $A^{\prime}.$  }
\end{enumerate}

} }

\end{enumerate}


The  new feature  of the original BMS group $B$,
as well as of all possible analogues of $B$ \cite{mac1}, the enlargement
of the finite dimensional vector space of translations
to the infinite dimensional vector space of supertranslations, is also shared by $B(2,1)$ and its
analogues \cite{Mel8}. In fact it is a common characteristic
of \it any \normalfont BMS group $\mathcal B$, and its analogues, in any number
$d$, $d\geq 3$, of space$-$time dimensions.
 This turns $\mathcal B$, as well as of all possible analogues of $\mathcal B$, in any number
$d$, $d\geq 3$, of space$-$time dimensions,
into infinite dimensional Lie groups.
Thus the required extension, which has been reduced above to the solution of three problems, is needed in order to construct
the IRS of all these groups.

In Subsections \ref{little} and \ref{solutions}
we find solutions to the three problems posed by the
required extension of Wigner$-$Mackey's theory.
These solutions apply both to the original BMS group $B$
and also to $B(2,1).$ They also apply, possibly with
appropriate modifications in each particular case,
to \it any \normalfont BMS group $\mathcal B$, and to \it any \normalfont of its analogues, in any number
$d$, $d\geq 3$, of space$-$time dimensions.
In Subsection
\ref{little}
we find a solution to the First problem
posed by the needed extension of
Wigner$-$Mackey's theory. This solution
applies to any topology $\mathcal B$ comes equipped with.
The solution to the Second problem, given in
Subsection \ref{solutions}, applies to the Hilbert
and to the nuclear topologies. Finally, the solution to the Third problem, given in
Subsection \ref{solutions}, applies to the Hilbert
 topology only. Thus solution to all three problems
 posed by  the
required extension of Wigner$-$Mackey's theory are
given in the Hilbert topology.


\vspace{0.5cm}
In the next Subsection we clarify which is the Hilbert topology the group $B(2,1)$
comes equipped with, and we point out
that $B(2,1)$ is not locally compact in this topology.

\subsection{
$B(2,1)$ \ is \ not \ locally \ compact \ in \ the \ Hilbert \ topology
}


\noindent
$L^{2}(P_{1}(R),\lambda ,R)$
is
topologised \cite{Melas} as a (pre) Hilbert space by using a natural measure $\lambda$
on $P_{1}(R)$ and by introducing
scalar product (\ref{ip}) into $ L^{2}(P_{1}(R),\lambda ,R).$
The norm defined by this scalar product  induces a metric in
$ L^{2}(P_{1}(R),\lambda ,R). $
$ L^{2}(P_{1}(R),\lambda ,R)$ is endowed
with the topology whose open sets are the
balls determined by this metric, the Hilbert
topology; the Hilbert topology is also called in the
literature metric, norm, or, strong topology.
If $ R^{4}$ is endowed with the natural metric topology then
the group $  G=SL(2,R),$ considered as
a subset of $ R^{4},$ inherits the induced topology on $G.$

We name the product topology of $
L^{2}(P_{1}(R),\lambda ,R)
\times   G,$ Hilbert topology,
to emphasize that $L^{2}(P_{1}(R),\lambda ,R)$
is endowed with the topology induced by the
scalar product (\ref{ip}) of the Hilbert space
$L^{2}(P_{1}(R),\lambda ,R).$
With the Hilbert topology, $B(2,1)$
becomes a topological group, in particular
a non$-$locally compact
group; the proof follows the corresponding proof
 \cite{Cantoni} for the original BMS group $B$.


\subsection{Orbits and ``little groups''}
\label{orbits}
Many symmetry groups  in Physics, which are isometry groups
of the underlying manifolds on which the dynamics unfolds,
 have the
structure of a semidirect product
\be
\label{sg}
\mathcal B = A  \rtimes_{T} \mathcal G,
\ee
where $A$ is a finite dimensional vector space, in particular $A=K^{d}$, $d\geq2$, $K$ is either the field $R$ of real numbers, or $C$  of complex numbers; thus
$A$ is the $K$$-$vector space of $d$$-$component column vectors with entries in $K$. $A$, being a vector space,
is an abelian group which is assumed to be a normal subgroup of $\mathcal B$. Now let $s: A \times A \longrightarrow K$ be any scalar product (i.e. any non$-$degenerate symmetric $K$$-$bilinerar form) for
$A$.
$\mathcal G$ is the identity component of the ``Lorentz group''  of linear transformations preserving $s$.
The remaining structure of the semidirect
product is specified by the action $T$ of
$\mathcal G$ on $A$ given by
\be
T(g)\alpha= g \cdot \alpha,
\ee
where $g \in \mathcal G, \ \alpha \in A,$ and $g \cdot \alpha$ is matrix multiplication of $g \in \mathcal G$ from  the left with $\alpha \in A$. Examples of symmetry groups with the semidirect product structure (\ref{sg})
 are the connected Euclidean group and the connected Poincare group in any number of dimensions $d$, $d\geq2$, and also, the
connected Ultrahyperbolic group in any number of dimensions $d=2n$, $n=2,3... \ .$

In order to find the IRS of these groups, by applying
Wigner$-$Mackey's theory, the first step is
to find the ``little groups''.  Determination
of the ``little groups'' in turn typically requires (see e.g. \cite{Barut}, p. 510, 516, 517)  classification of the $\mathcal G$ orbits on the dual group $\hat{A}$ of $A$, or equivalently,
due to the bijection (\ref{bc}),
 classification of the $\mathcal G$ orbits on the topological dual  $A^{\prime}$ of $A$.
The key fact is that $A$ is a \it{finite} \normalfont dimensional vector space and
consequently its topological dual $A^{\prime}$ is a \it{finite} \normalfont dimensional vector space isomorphic to $A^{\prime}$,
and therefore, of
the same dimension as $A$. This makes
classification of the $\mathcal G$ orbits
on $A^{\prime}$ a fairly easy task (see e.g. \cite{Barut}, p. 516). In all cases the $\mathcal G$ orbits
are parameterized by  finite sets of parameters.

However, in the case of
$B(2,1)$, $\mathcal G=G=
S
L
(2,R)$, and $A= {L^{2}}(P_{1}(R),\lambda ,R),$  which is an \it{infinite}
\normalfont
dimensional Hilbert  space.
As we pointed out in Subsection \ref{sec2.3},
Reisz$-$Fr$\acute{e}$chet theorem for Hilbert spaces
states that
the topological dual $A^{\prime}$ of $A$, is isomorphic to $A$.
Therefore, $A^{\prime}\simeq {L^{2}}(P_{1}(R),\lambda ,R)$, and
$A^{\prime}$ is itself an \it{infinite}
\normalfont
dimensional Hilbert  space.
Hereafter we shall identify $A^{\prime}$ with
 ${L^{2}}(P_{1}(R),\lambda ,R)$ and we shall
 write $A^{\prime}= {L^{2}}(P_{1}(R),\lambda ,R),$
 with the understanding that $A^{\prime}$ and
 $L^{2}(P_{1}(R),\lambda ,R)$ are in fact isomorphic.

 Infinite dimensional Hilbert  spaces are too spacious
to allow, in general, an easy classification of the
orbits of a group action
on them.
When $A^{\prime}$ is an
infinite dimensional Hilbert  space
the orbits of the $\mathcal G$ action
on it, in general,
are parameterized by countably infinite sets of parameters, and only a partial
classification of them is feasible (see e.g. \cite{mac4,Melas}).
This is
precisely what happens \cite{Melas} in the case of $B(2,1)$, classification of the $
S
L
(2,R)$ orbits on $A^{\prime}={L^{2}}(P_{1}(R),\lambda ,R)$ can only be partially
achieved.

To conclude, determining  the ``little groups'' by classifying
the $\mathcal G$ orbits is appropriate when $A$ is
a finite dimensional vector space; this
is precisely what Wigner and Mackey did in their
initial considerations \cite{Wigner,Mackey0,Mackey}.
On the other hand, when
$A$ is an infinite dimensional vector space,
and only, in general, a partial classification of the
$\mathcal G$ orbits can be attained, determining  the ``little groups'' by classifying
the $\mathcal G$ orbits is not appropriate,
and a different method should be followed. This is
precisely the problem which is addressed and solved
by the \it{Programme} \normalfont put forward in the
next Subsection.

\subsection{Determination of the ``little groups''
}
\label{little}
Hereafter, we restrict attention to semidirect products
$\mathcal B = A  \rtimes_{T} \mathcal G$
which share the new property of $B(2,1)$, namely,
their largest proper
normal abelian subgroup $A$, is also an
infinite dimensional vector space. In fact it is assumed that  $A$
is an infinite dimensional Hilbert space of real$-$valued
functions defined on a locally compact space $\mathcal Q$,
square integrable with respect to $\lambda$, $\lambda$ being a natural measure on $\mathcal Q$.

The subgroup $A$ may be topologized
as a (pre) Hilbert space by introducing a scalar product into $A$        and by using a natural measure $\lambda$ on $\mathcal Q$.
In the product topology of $A \times \mathcal G$, $\mathcal B$
becomes a topological group. In the case of $B(2,1)$, $A$ is topologized as a (pre)Hilbert space by introducing into $A$ the scalar product
\be
\label{ip}
<\alpha(\theta),\beta(\theta)>=\int_{P_{1}(R)}\alpha(\theta) \beta(\theta) {\rm d} \lambda(\theta)
\ee
where $\lambda$ is chosen to be the standard
normalized Lesbegue measure $ {\rm d} \lambda = \frac{{\rm d} \theta}{2 \pi}   $ on the circle.

As the  discussion in the previous Subsection suggests, in order to find the ``little groups'' when $A$ is an infinite dimensional vector space, we need to follow
another route \cite{mac6}. Firstly a definition is in order:

\begin{definition}
For each closed subgroup $\mathcal S$ of $\mathcal G$, define
its maximal invariant  vector subspace $\mathcal V$($\mathcal S$)
of $A^{\prime}$ by
\be
\mathcal V(\mathcal S)= \{ f \in A^{\prime}| \ s f = f \ \ \rm for \ \rm all \ s \in \mathcal S \}. \ee
\end{definition}
$\mathcal V$($\mathcal S)$ is closed since the $\mathcal G$ action on $A^{\prime}$
is continuous. In the case of $B(2,1)$, $A^{\prime}={L^{2}}(P_{1}(R),\lambda ,R),$
and consequently, $\mathcal V$($\mathcal S)$ being a closed subspace of
${L^{2}}(P_{1}(R),\lambda ,R),$ is itself a Hilbert space.

Now the following holds \cite{mac6}:
\begin{theorem}
Every ``little group'' of $\mathcal B$ is a closed Lie subgroup of $\mathcal G$. The ``little groups'' fall into three classes:
\begin{enumerate}
\item{Discrete subgroups,}
\item{Connected Lie subgroups,}
\item{Non$-$connected  non$-$discrete Lie subgroups.}
\end{enumerate}
\end{theorem}
The identity subgroup $e \in \mathcal G$  belongs both to 1 and 2,
otherwise the three classes are non$-$overlapping.

When $A$ is infinite dimensional, the ``little groups'' of $\mathcal B$ may  be found, in principle, by following the \it{Programme} proposed by McCarthy
\cite{mac6}.

\vspace{1mm}

\noindent
\it{Programme} \normalfont

\begin{enumerate}
\item{Determine, up to conjugacy, all subgroups $
\mathcal S \subset \mathcal G$ belonging to the three classes 1, 2, and 3.}
\item{For each such $\mathcal S$ find the vector space $\mathcal V (\mathcal S)$.}
\item{Find the subgroups $\mathcal S$ for which $\mathcal V (\mathcal S)$ contains elements fixed under no properly larger group $\mathcal S^{'} \supset \mathcal S $. These last subgroups, and only these, are the ``little groups''.}
\end{enumerate}

\noindent
The following remarks are now in order:
\begin{enumerate}
\item
{Regarding the first part of the \it{Programme} \normalfont we note that as it
was
pointed out
in the previous Subsection, it is not appropriate, when $A$ is an infinite dimensional vector space,  to determine  the ``little groups'' by classifying
the $\mathcal G$ orbits. Instead, since
``little groups'' are subgroups of
$\mathcal G$, in order to find
the ``little groups'', we start by finding, up to
conjugacy, \it{all} \normalfont subgroups $\mathcal S$ of $\mathcal G$ belonging to the three classes 1, 2, and 3. Up to conjugacy, because conjugate ``little groups'' lead to equivalent IRS of of $\mathcal G.$ Finding the
subgroups $\mathcal S$ of $\mathcal G$ is a problem on its own right \cite{mac3}, which occasionally may be hard to solve \cite{Mel5},
or even does not have a complete solution \cite{mac6} with our
current knowledge of affairs.}

\item
{Regarding the second part of the \it{Programme} \normalfont we note that for \it{each} \normalfont subgroup $\mathcal S$ of $\mathcal G$, belonging to the three classes 1, 2, and 3, we search for non$-$zero
$f  \in A^{\prime}$ which are invariant under $\mathcal S.$ If there are such non$-$zero
$f  \in A^{\prime}$ which are invariant under  a subgroup $\mathcal S$ of
$\mathcal G,$  then the set $\mathcal V(\mathcal S)$ of
all such non$-$zero $f$, as we already pointed out, is a
vector space, in fact it is a Hilbert space, subspace of
$A^{\prime}={L^{2}}(P_{1}(R),\lambda ,R).$ It is appropriate now to draw attention to the following points:
\begin{enumerate}
\item {An element  $f$ is  non$-$zero as an element of the Hilbert space $A^{\prime}.$}

\item{It is easy to show that if $f_{1}$ and $f_{2}$
are two distinct elements of  $\mathcal V(\mathcal S)$ then the
orbits $\mathcal G f_{1}$ and $\mathcal G f_{2}$ are
also distinct.}

\item{If a subgroup $\mathcal S$ of $\mathcal G$ has non$-$zero invariant functionals $f  \in A^{\prime},$ then such a subgroup $\mathcal S$ \it does not \normalfont qualify yet as ``little group.'' Such a
    subgroup $\mathcal S$ is a \it potential \normalfont ''little group''.}
\end{enumerate}

\hspace{0.4cm} $\mathcal S$ is a \it potential \normalfont ''little group''
because \it a priori \normalfont we cannot exclude the
possibility that \it all \normalfont the elements
$f  \in \mathcal V(\mathcal S)$  are also invariant under some bigger
group $\mathcal S^{*},$ where, $\mathcal S^{*} \subseteq \mathcal G.$ If this were the case then $\mathcal S$ would
not qualify as an actual ''little group''. On the other hand,
if for some group $\mathcal S,$ there is at least one
element
$f \in \mathcal V(\mathcal S),$ which is invariant only under  $\mathcal S,$ and which has no higher invariance, then the potential ''little group'' $\mathcal S $ does qualify as an actual ''little group''.}

\item {Regarding the third part of the \it{Programme} \normalfont we note that, as we explained before, in order to find the (actual) ''little groups'' we need to find those subgroups  $\mathcal S$ of $\mathcal G$ which have at least one
    invariant functional $f \in \mathcal V(\mathcal S)$
    with no higher invariance. This is a problem on its own right. Usually, after having identified $\mathcal V(\mathcal S),$  one proceeds with brutal force \cite{macMel,Mel6,Melas} and constructs explicitly such an $f$; a task which is by
    no means trivial and can take considerable amount of time and effort \cite{macMel,Mel6,Melas}. If such an $f \in \mathcal V(\mathcal S)$ is found then the potential ''little group''   $\mathcal S $ does qualify as an actual ''little group''.}
\end{enumerate}

The implementation of each part of the Programme is by no means trivial, and it poses its own problems and challenges \cite{mac3,mac4,mac6,Mel5,Mel6,Mel7,Melas}. In the case of $B(2,1)$ firstly we prove \cite{Melas} that the ``little groups'' are compact in the employed Hilbert topology. Knowing that the ``little groups'' are compact
facilitates subsequently the implementation of all parts of the Programme because
search now is restricted  to the compact subgroups $\mathcal S$ of $\mathcal G$ which belong to the three classes 1, 2, and 3. Moreover, the vector space $\mathcal V (\mathcal S)$ is calculated only for the compact subgroups $\mathcal S$ of $\mathcal G$, and only among these groups we isolate those which contain elements fixed under no properly larger
group.

\subsection{
Extension of Wigner$-$Mackey's theory
accomplished in the Hilbert topology
}
\label{solutions}
As we pointed out in
Subsection \ref{extension}
Wigner$-$Mackey's theory
needs to be extended to be
applied to semidirect products
$\mathcal B = A \rtimes_{T} \mathcal G,$
when the Abelian group $A$
is also an infinite dimensional vector space. In order to be applied to this class of semidirect products Wigner$-$Mackey's theory needs to be extended in such a way
that solves the three problems stated in
Subsection \ref{extension}:
The ``little groups'' problem, the irreducibility problem, and the exhaustivity problem.

We note that Wigner$-$Mackey's theory is an enormously effective, systematic tool for
analyzing representations of many different groups,  has been used to good effect
by many workers, and has been extended in different directions over the years.
The extension given here is the appropriate  one in order to
study and construct the IRS of the BMS group $B$, and the IRS of its analogues,
in any number $d$
of space$-$time dimensions, $d\geq3$, and also in order to construct the IRS
of their supersymmetric counterparts.

In Subsection
\ref{little}
we solved
the ``little groups'' problem.
In order to solve the problem
we restricted attention to semidirect products
$\mathcal B = A \rtimes_{T} \mathcal G,$
where $A={L^{2}}(\mathcal Q,\lambda ,R)$
is a separable infinite dimensional Hilbert space of real$-$valued
functions defined on a locally compact space $\mathcal Q$,
square integrable with respect to $\lambda$, $\lambda$ being a natural measure on $\mathcal Q$.

This is a large class of
semidirect products which encompasses both the original BMS
group $B$ \cite{mac3,mac8,mac4,mac5,mac6} and its generalizations \cite{mac1} and the group $B(2,1)$ 
and its generalizations \cite{Mel8}, when the functions which inhabit
$A$ are  square integrable with respect to the appropriate
measure in each case.

For this class of semidirect products,
when $A$ is endowed with the Hilbert topology, we can also
solve the other two problems: The irreducibility problem, and the exhaustivity problem.  It is noteworthy
that the   solution to the ``little groups'' problem,
given in Section \ref{little},
applies  also to all
``reasonable'' topologies
$A$ can be endowed with.
These topologies
 are explicated in the next section.

  To conclude solutions to the
  three problems, stated in Section \ref{extension},
when
$\mathcal B$ is employed with
the Hilbert topology,
are obtained as follows:
\begin{enumerate}
\vspace{0.2cm}
\item{\it{First problem.
The ``little groups'' problem: }
\normalfont
This problem was
solved in
Subsection
\ref{little}.
}

\vspace{0.2cm}
\item{\it{Second problem.
The irreducibility problem: }
\normalfont When $\mathcal B$ is not locally compact Mackey's theorems 14.1
and 14.2 in \cite{Mackey0},
which prove the irreducibility of the representations of $\mathcal B$ derived by the inducing construction,
do not no longer apply. However, when $\mathcal B$
is employed with the Hilbert or nuclear topology,
by using the results
of \cite{mac5} and  \cite{mac6} respectively,
we can prove that the representations of $\mathcal B$
obtained by Wigner$-$Mackey's inducing construction are
irreducible despite the fact that $\mathcal B$  is not
locally compact in the employed Hilbert or nuclear topology. In the case of $B(2,1)$, and
for the Hilbert topology,  we give the proof in \cite{Melas}.
}


\vspace{0.2cm}
\item{\it{Third problem.
The exhaustivity problem: \normalfont
As we pointed out in Subsection \ref{extension},
when $A$ is an infinite dimensional  vector space, and
hence
$\mathcal B$ is not locally compact,
Mackey's classic regularity condition, i.e.,
the existence of a Borel cross section
 which meets each $ \mathcal G-$orbit in
  $ A^{\prime} $ exactly once, which
ensures the exhaustivity of Wigner$-$Mackey's
inducing construction,
cannot be used
  as a sufficient condition,  and
 new
 sufficient conditions
which insure
the exhaustivity of the IRS
of $\mathcal B$ constructed
by Wigner$-$Mackey's
``little group'' method
  need to be introduced.

\hspace{0.4cm}  This was done in
 two remarkable papers by Piard \cite{Piard1,Piard2} in the case where
 $\mathcal B$ is endowed with the Hilbert
 topology. Piard followed what Mackey did \cite{Mackey0,Mackey00},
when $A$ is a finite dimensional vector space, and he
appropriately adapted and modified it to the case
where $A$ is an infinite dimensional vector space.

\hspace{0.4cm} The sufficient conditions he
 introduced entail the strong topology
 of $A$ and the weak topology of $A^{\prime}$.
Piard  proved  \cite{Piard2} that the original
BMS group $B$ satisfies the sufficient conditions he introduced and so he proved
\it that in the case of $B$ Wigner$-$Mackey's
``little group'' method gives all the IRS of
$B.$ \normalfont This proves in particular
that McCarthy in \cite{mac3} and \cite{mac4} obtained all the IRS of $B$
and this renders his results even more
important.

\hspace{0.4cm} It is not a trivial task
to check if a BMS group $B$ in any number
of $d\geq3$ space$-$time dimensions
satisfies Piard's sufficient conditions.
We undertake this task in \cite{Melas} and
we prove that the group $B(2,1)$ satisfies
Piard's sufficient conditions and therefore
the IRS of $B(2,1)$ obtained here by
applying Wigner$-$Mackey's
``little group'' method are exhaustive.

\hspace{0.4cm}The first one
third of Piard's paper deals also with the case
where $A$ is endowed with the nuclear topology.
It is an interesting and important open  problem
to complete Piard's investigation, or apply
a totally different approach, and prove or
disprove that we obtain all IRS of $\mathcal B$,
by applying Wigner$-$Mackey's inducing construction, when $A$ is equipped with the
nuclear topology.

\hspace{0.4cm}This will inform us
in particular if McCarthy  in \cite{mac6}, where he
equipped $A$ with the nuclear topology, obtained all IRS of the original BMS group $B$, and moreover, if we obtain all IRS of $B(2,1)$
by applying Wigner$-$Mackey's inducing construction, if we endow $A$ with the
nuclear topology. Similar remarks apply to
any  any BMS group $B$ in any number
of $d\geq3$ space$-$time dimensions.

}}
\end{enumerate}





\vspace{0.4cm}

It is possible and desirable,
as it is explained in the next section,
to endow $B(2,1)$ with other topologies, different from the Hilbert
topology. It is not known if $B(2,1)$ is locally compact
in these other topologies.
Therefore,
if $B(2,1)$ is not locally compact in these other topologies
we should
examine anew
if
the representations of $B(2,1)$ we obtain with Wigner$-$Mackey's inducing construction are irreducible, and, if we obtain all of them.

\section{Other choices of topologies}
\label{top}

It is amusing that one can understand intuitively
\cite{mac8,Melas} that the ``little groups'' are compact in terms of
light intensity distributions on distant ``celestial'' spheres.
The dual action $T'$ of $ G$ on $ {L^{2}}^{'}(P_{1}(R),\lambda ,R)$
is given by equation (\ref{tzatziki}):
$$
(T'(g)f)(\sigma)=\kappa_{g}^{-2}(\sigma)f(\sigma g).
$$

This action is similar to the transformation law for
light intensity distributions on distant (``celestial'') 1$-$dim   spheres, i.e. circles, under $G$. The latter action appears with a $\kappa_{g}^{-1}$ rather than $\kappa_{g}^{-2}$ factor (see e.g. \cite{Greiner}, p.405, the discussion there refers to  4$-$dim space$-$time but it is easily
adapted to the case of 3$-$dim space$-$time); but the analogy works nevertheless. The $\kappa_{g}$ factor corresponds physically to a
blueshift factor. A subgroup $\mathcal S$ of $ G$ is a ''little group'' if it leaves a (non$-$zero) $f$ fixed under $T'.$  We showed
before that $ \mathcal S$ is closed. Consequently, if $ \mathcal S$
is non$-$compact, it must contain a sequence of pure boosts whose velocity parameter becomes arbitrary close to the speed of light.
For a light intensity distribution $f$ to be invariant under $\mathcal S$,  it must be such that, when subjected to arbitrary large
blueshifts over a region of the circle with arbitrary small complement
(the complement is redshifted), it remains the same. It is not surprising that there are no such non$-$zero $f's$, so that there cannot be any
non$-$compact ``little groups''.

In this paper we restrict attention to the case where
the supertranslation space $F(P(N^{+}))$
is
realized as
a separable Hilbert space, namely  $ L^{2}(P_{1}(R),\lambda ,R).$
We equip $ L^{2}(P_{1}(R),\lambda ,R)$ with the Hilbert
topology and topologize $B(2,1)$ with the product topology of
$L^{2}(P_{1}(R),\lambda ,R)\times   G.$
There is a wide range  \cite{Crampin2} of
``reasonable'' topologies by which we can endow $B(2,1).$
This freedom we have in the choice of the topology for $B(2,1)$
stems
from the infinite$-$dimensional additive supertranslation
subgroup $F(P(N^{+}))$ of ``arbitrary'' real$-$valued functions
defined on the $P(N^{+}) \simeq S^{1}$.
The range of
choices available depends on the class of functions
allowed in $F(P(N^{+})).$

Regarding the class of functions allowed in $F(P(N^{+}))$
we note the following: In his original derivation of the
BMS group $B$ Sachs took supertranslations \cite{Sachs1}
to be twice continuously differentiable functions, since
they are coordinate transformations. Sach's original derivation of $B$ was superseded by that of Penrose \cite{Pen1,Pen2,Pen3},
who gave a precise and intrinsic derivation of $B$ as that group
of \it exact \normalfont conformal motions of the future
(or past) null boundary $\Im^{+}$ (or $\Im^{-}$) of
(conformally compactified weakly asymptotically simple)          space$-$times which preserve ``null angles''. Since truly arbitrary supertranslation functions describe symmetry transformations
in Penrose's sense, supertranslations need not have some minimum degree of smoothness.

$B(2,1)$ can also be derived in Penrose's sense, i.e., as as that group
of \it exact \normalfont conformal motions of the future
(or past) null boundary $\Im^{+}$ (or $\Im^{-}$) of 3$-$dim
(conformally compactified weakly asymptotically simple)          space$-$times which preserve ``null angles''. Consequently
similar remarks for the available freedom in the choice of
the degree of smoothness of the supertranslations apply also in
the case of $B(2,1).$

\indent
We pointed out before that
the dual action $T'$ of $ G$ on $ {L^{2}}^{'}(P_{1}(R),\lambda ,R)$
 is similar to the transformation law for light intensity distributions on distant (``celestial'') 1$-$dim   spheres under $G$, and this suggests, that there cannot exist any invariant,
 under non$-$compact  ``little group'' $\mathcal S$,
 square integrable $f \in {L^{2}}^{'}(P_{1}(R),\lambda ,R).$
However, the same reasoning gives rise to the possibility
that there may exist invariant
\it distributional \normalfont $f$ invariant under
non$-$compact  ``little groups''.

\indent
Indeed this is what happens in the 4$-$dim case \cite{mac6}, when $F(P(N^{+})),$ $P(N^{+}) \simeq S^{2}$ in the 4$-$dim case,
is chosen to be the space $C^{\infty}(S^{2})$ of infinitely differential functions on the 2$-$dim sphere. If a sufficiently fine (actually nuclear) topology is chosen for $C^{\infty}(S^{2})$ then non$-$compact  ``little groups'' arise associated with invariant distributions. We conjecture that the same is going to happen in the 3$-$dim case: If $F(P(N^{+}))$
is chosen to be the space $C^{\infty}(S^{1})$
of infinitely differential functions on the circle,
and equip this space with the nuclear topology then
non$-$compact  ``little groups'' are going to arise associated with invariant distributions.

\indent
IRS both in the Hilbert topology and in the nuclear topology induced by compact ``little groups'' describe bounded sources, whereas,  IRS in the nuclear topology induced by non$-$compact
``little groups''  describe scattering states \cite{mac6,Crampin2}. Another class of topologies for
$F(P(N^{+}))$ are the Sobolev topologies \cite{Crampin2}, which depend strongly on smoothness
properties (being defined by convergence in the mean of functions and derivatives). They are physically plausible in a gravitational  context because they  are adapted to initial$-$value problems for hyperbolic differential equations; in an ADM formulation of G.R. the evolution equations of the metric and the extrinsic curvature are hyperbolic equations.







\section{Construction of the IRS of $B(2,1)$}

\label{sec4}
\subsection{Prerequisites}
\noindent
Let
$\mathcal B = A  \rtimes_{T} \mathcal G $
be the semidirect product of $A$ and $ \mathcal G,$
specified by the continuous action $T: \mathcal G
\longrightarrow  {\rm Aut} (A).$
In section \ref{sec2} we pointed out that there is a
bijective correspondence between the dual group $\hat{A}$ of $A$ and the topological dual $ A^{\prime}$ of $ A$.
This bijection is also
useful here so we recall the explicit form of the
bijection between $\hat{A}$  and the  $ A^{\prime}$ and draw a conclusion not stated in Section \ref{sec2}.
 $\hat{A}$  and  $ A^{\prime}$ are isomorphic
as real vector spaces, and in particular as abelian groups. The isomorphism $j$ between  $\hat{A}$  and  $ A^{\prime}$ is given by equation
(\ref{isoj}):
$$
j:  A^{\prime}  \rightarrow  \hat{ A}, \ \ \
j(f)  =  e^{i f}. \nonumber
$$

Therefore every element $j(f)=\phi \in \hat{ A}$
is written
as $\phi=e^{i f}$, where
$f$ is determined uniquely by $\phi$
and vice versa. We have (equation (\ref{chi})):
$$
\phi(\alpha)=e^{i f(\alpha)},
$$
$\alpha \in A.$
The action $T$ of $\mathcal G$ on $\hat{A}$
is defined by (equation (\ref{indact})):
$$
(b \phi)(\alpha)=\phi(b^{-1}\alpha),
$$
$b \in \mathcal G, \ \phi \in \hat{A}, \ \alpha \in A.$ The bijective correspondence
$
\hat{ A} \leftrightarrow  A^{\prime}
$
is preserved pointwise by the $\mathcal G$ action
if and only if the action of $\mathcal G$
on $ A^{\prime}$ is defined by (equation (\ref{da})):
$$
(gf)(\alpha)=f(g^{-1}\alpha),
$$
$g \in \mathcal G, \ f \in A^{\prime}, \ \alpha \in A.$

In Section \ref{sec2} we defined the ``little group''
$L_{\phi} $ of $ \phi,$ $\phi \in \hat{A}, $
to be the largest subgroup
of $\mathcal G $ which leaves $\phi $ fixed.
Similarly, we can define the ``little group''
$L_{f} $ of $ f,$ $f \in A^{\prime}, $
to be the largest subgroup
of $\mathcal G $ which leaves $f $ fixed.
It is an immediate consequence of equations
(\ref{chi}), (\ref{indact}), and  (\ref{da}),
that if $f$ is the unique element of  $A^{\prime}$
which corresponds to the element $\phi $ of
$ \hat{A}, $ then, $L_{f}=L_{\phi},$ i.e., we have:
\be
\label{iso}
\rm {If}  \ \ \it{f} \longleftrightarrow \phi \ \ \rm {then} \ \
\it L_{f}=L_{\phi}.
\ee

The last equality,  not mentioned in Section \ref{sec2},
is going to be useful in the subsequent construction of the IRS of $B(2,1)$.
One more remark is now in order before proceeding to the construction of the
IRS of $B(2,1)$.
As we  stated in Section \ref{sec2},
the set of characters which can be
reached from a character $ \phi $ by the
$  \mathcal G $ action on $\hat{A}$ (equation (\ref{indact}))   is called the orbit
of $ \phi $, and is denoted by $ \mathcal G \phi.$
There is a bijection between the orbit $ \mathcal G \phi$
and the coset space $\mathcal G / L_{\phi}$,
the set of left cosets of $L_{\phi}$ in $\mathcal G,$
(equation (\ref{omg}))
\be
\label{bij1}
 \mathcal G \phi \longleftrightarrow \mathcal G/L_{\phi}.
\ee

Similarly, the set of functionals which can be
reached from a functional $ f  $ by the
$  \mathcal G $ action on $A^{\prime}$ (equation ((\ref{da}))   is called the orbit
of $ f  $, and is denoted by $ \mathcal G f .$
There is a bijection between the orbit $ \mathcal G f $
and the coset space $\mathcal G / L_{f }$,
the set of left cosets of $L_{f }$ in $\mathcal G,$
\be
\label{bij2}
 \mathcal G f \longleftrightarrow \mathcal G/L_{f}.
\ee
Both bijections (\ref{bij1}) and (\ref{bij2})   will even be homeomorphisms \cite{Bour} (Appendix I), under quite general conditions.

From equations (\ref{iso}), (\ref{bij1}), and (\ref{bij2}), we conclude that if
$\phi$ is the unique element of  $\hat{A}$
which corresponds to the element $f $ of
$ A^{\prime}, $ then, there is a bijective correspondence between the orbits $\mathcal G \phi$ and
$\mathcal G f$, i.e.,  we have
\be
\label{bij3}
\mathcal G \phi  \longleftrightarrow \mathcal G f.
\ee
Therefore we can identify, if needed, the sets
$\mathcal G \phi$ and $G f$.
The previous comments also apply to $B(2,1)$, in which case we have,
 $A^{\prime} \simeq A= {L^{2}}(P_{1}(R),\lambda ,R)$, and,    $\mathcal G=G=
S
L
(2,R)$.

\subsection{Necessary data for the construction of the IRS}

To find explicitly the operators of the induced
representations of $B(2,1)$, it suffices to provide the information
stated in paragraphs 1 and 2 at the end of  Section \ref{sec2} for each of the orbit types.
It is convenient, in order to construct the operators of the IRS of
$B(2,1)$, to use the bijective correspondence between the
 dual group $\hat{A}$ of $A$ and the topological dual $ A^{\prime}$ of $ A$
 and the consequences of this correspondence, i.e., to use equations
(\ref{isoj}), (\ref{iso}), and (\ref{bij3}),  and restate the
 information
stated in paragraphs 1 and 2 at the end of  Section \ref{sec2} as follows:
In order to give
the operators of the IRS of $B(2,1)$  explicitly
it is
necessary to give the following information
\begin{enumerate}
\item{
An irreducible unitary representation $U$ of $L_{f}$ on a
Hilbert space $D$ for each $L_{f}.$}
\item{
A $ \mathcal G$$-$quasi$-$invariant measure $ \mu $
on each orbit $ \mathcal G f
\approx \mathcal G / L_{f};$
where $L_{f} $ denotes the ''little group'' of the base point
$ f \in A^{\prime} $ of the orbit
$ \mathcal G f.$
}
\end{enumerate}

Thus for \it each \normalfont orbit
$ G f \in A^{\prime}$, with
base point
$f,$ and \it each \normalfont irreducible unitary representation $U$ of $L_{f}$, we have
an irreducible representation of
$B(2,1),$ given by equation
(\ref{ukygvkhvkyutfkyufvmhgv}), which is
unitary, provided the orbit $ G f$ comes equipped with a measure which
is $ G$$-$quasi$-$invariant.
We conclude that in order to construct
the operators of the IRS of $B(2,1)$
we need to give the following data
in the prescribed order:
\begin{enumerate}
\item{``Little groups'' and their invariant functionals.}
\item{IRS of the ``little groups'' on a Hilbert space $D$.}
\item{A $ G$$-$quasi$-$invariant measure
on each orbit $  G f$.
}
\end{enumerate}
We proceed now to carry out this task.


\subsection{``Little groups'' and their invariant functionals}

\noindent
The calculation of the ``little groups'' goes
hand in hand with the calculation of their
invariant functionals, and it is achieved
by following the \it{Programme} \normalfont
stated in Subsection \ref{little}.

It is an important result
that when $B(2,1)$ is employed with the Hilbert topology \it {all} \normalfont ``little groups'' of $ B(2,1)$ are compact.
A formal proof is given in \cite{Melas}.
A heuristic, intuitive proof was given in Section
\ref{top}.

The maximal compact subgroup of $G$, where $G=SL(2,R)$, is
just the subgroup $SO(2).$ That is to say, if $K$ is a compact
subgroup of $G,$ then some conjugate $g K g^{-1}$ of $K$ is
a compact subgroup of $SO(2).$ In the representation theory
of $B(2,1)$, the little groups are only significant up to conjugacy. So, after a possible conjugation, we may take
every little group to be  a compact (or, equivalently, closed)
subgroup of $SO(2).$

This makes the Theorem which follows more intelligible.
This Theorem, proved in \cite{Melas},
describes in detail the  ``little groups'' of $ B(2,1).$
\begin{theorem}
\label{Lit}
The ``little groups'' $\;L_{f}\;$ for $B(2,1)$ are precisely the closed subgroups of $
S O (2)$ which contain the element $-I$, $I$ being the identity element of $SL(2,R)$. These are (A) $SO(2)$ itself, and (B)  the cyclic groups $C_{n}$ of
even order $n$.
\end{theorem}

The proof of the preceding Theorem involves identifying
$SO(2)$, and $C_{n},$ with $n$ even, as potential
``little groups'' and then proving that they do in fact
occur as actual ``little groups''. It is in this second part
of the proof that we proceed with brutal force
and actually construct in each case an invariant
functional with no higher invariance. This
construction, as we pointed out in Subsection
\ref{little}, is by no means a trivial task,
and can take in general \cite{macMel,Mel6}, and in fact it does so here,
considerable amount of time and effort,
at least as much as required to identify the
potential ``little groups'' at the first place.


As we pointed out in Section \ref{sec3}, for a given ``little group'' $\;L_{f}\;$ the elements $f \in A^{'}$ which are invariant under $\;L_{f}\;$, i.e., the
elements $f \in A^{'}$ which satisfy
\begin{equation}
(T'(g)f)(\theta)= f(\theta),
\end{equation}
$\theta \in P_{1}(R),$
form a (Hilbert) subspace of ${L^{2}}(P_{1}(R),\lambda ,R)$. We denote this subspace by $L^{2}(L_{f})$.
We have the following Theorem \cite{Melas}:
\begin{theorem}
The Hilbert space $L^{2}(S O (2))$ of invariant vectors $f \in A^{'}$ under $S O (2)$ is:
\begin{equation}
L^{2}(S O (2))=\left \{ f \in L^{2}(P_{1}(R),\lambda ,R) \; | \;  f(\theta)=c, \; c \in R \right \}.
\end{equation}
So $L^{2}(S O (2))$ is just the one$-$dimensional space of constant real$-$valued functions defined on $P_{1}(R)$.
The Hilbert space $L^{2}(C_{n})$ of invariant vectors $f \in A^{'}$ under $C_{n}$ is:
\begin{equation}
L^{2}(C_{n})=L^{2}(E_{n}),
\end{equation}
where $L^{2}(E_{n})$ is the Hilbert space of square integrable real$-$valued functions defined on
\begin{equation}
E_{n}= \left \{ \theta \in  P_{1}(R)  \; | \;    0   <  \theta  < \frac{4 \pi}{n} \right \}.
\end{equation}
\end{theorem}
It is important to bear in mind that, as we emphasized
in Section \ref{sec3},  if $f_{1}$ and $f_{2}$ are two
distinct elements of $L^{2}(S O (2))$, or of $L^{2}(C_{n})$,
then the orbits $ G f_{1}$ and $ G f_{2}$,
are also distinct.

\it{Thus each $f \in L^{2}(S O (2)),$
and, each $f \in L^{2}(C_{n}),$ gives rise to a different
orbit $G f,$  and on each such orbit a continuous irreducible
unitary representation of $B(2,1)$ can be built in the
way summarized in paragraphs 1 and 2 of the previous Subsection. Representations built on distinct orbits are also
distinct, i.e., not unitary equivalent. }
\normalfont

Moreover, in  \cite{Melas} it is shown that the Wigner$-$Mackey's inducing construction is exhaustive
despite the fact that $ B(2,1)$  is not
locally compact in the employed Hilbert topology.
This result is rather important because other group theoretical approaches
to quantum gravity which invoke Wigner$-$Mackey's inducing construction (see e.g. \cite{IshKak0,IshKak}) are typically plagued
by the non$-$exhau$-$ \newline stiveness of the inducing construction which results precisely from the fact that
the group in question  is not locally compact in the prescribed topology.
Exhaustiveness is not just a mathematical nicety:
If the inducing construction is not exhaustive one cannot
know if the most interesting information or part of it
is coded in the irreducibles which cannot be found by the Wigner$-$Mackey's inducing procedure.
These results, i.e.
compactness of the ``little groups'' and exhaustiveness of the inducing construction, not only are
 significant  for the group theoretical approach to quantum gravity advocated here, but also  they have
 repercussions \cite{Melas}  for  other approaches to quantum gravity.

\subsection{IRS of the ``little groups'' on a Hilbert space $D$.}
\noindent
The ``little groups'' of $B(2,1)$ are given in Theorem \ref{Lit} and are all abelian.
All the irreducible
representations of an
 abelian group are one$-$dimensional. We have in particular:
\newline

\noindent
\bf (1).
\normalfont     \hspace{0.3cm} $L_{f}=SO(2)$.
\newline
\noindent
The IRS $U$ of $SO(2)$ are parameterized by an integer $\nu$ which for distinct representations takes  the values $\nu =...,-2,-1,0,1,2,... \ .$
Denoting these representations by $U^{(\nu)}$,
they are given by multiplication in one complex dimension
$D \approx C $ by
\begin{equation}
U^{(\nu)}(R(\theta))=e^{i \nu \theta},
\end{equation}
where $ R(\theta)=  \begin{pmatrix}
    cos\theta & -sin\theta \\
    sin\theta & cos\theta
  \end{pmatrix} \in SO(2)$.

\noindent
\bf (2).
\normalfont     \hspace{0.3cm}
$L_{f}=C_{n}$, $n$ is even.
\newline
\noindent
The IRS $U$ of $C_{n}$ are parameterized by an integer $\lambda$ which for distinct representations takes  the values $\lambda =0,1,2,...,n-1.$
Denoting these representations by $U^{(\lambda)}$,
they are given by multiplication in one complex dimension
$D \approx C $ by
\begin{equation}
U^{(\lambda)} \left (R \left ( \frac{2 \pi}{n}j \right ) \right )=e^{i  \frac{2 \pi}{n} \lambda j},
\end{equation}
where $ j$  parameterizes the elements of the group $C_{n}$.

\subsection{ A $ G$$-$quasi$-$invariant measure on each orbit $Gf$.}
\label{invmeas}


\noindent
 A $\;  G\;$$-$quasi$-$invariant
measure $\; \mu \;$ on each orbit $\;  G f \approx
 G / L_{f}\;$ is required. However, a $\;  G\;$$-$invariant
measure $\; \mu \;$ on each orbit $\;  G f \approx
 G / L_{f} \;$
can
be given in all cases.

\subsubsection{Invariance and quasi$-$invariance}

\noindent
It is useful at this point to recall some basic facts
from elementary measure theory \cite{Bourbaki}.
A measure $\mu$ on a measurable space $\mathcal M$ is called
 $\mathcal G$$-$invariant, or invariant with respect to
$\mathcal G$, if for any measurable set $E \subset \mathcal M$,
and $k \in \mathcal G$, the transformed measure $k \mu$, also
denoted by $\mu_{k}$, defined by
\be
\mu_{k}(E)=\mu(k^{-1}E),
\ee
satisfies the relation
\be
\mu_{k}(E)=\mu(E),
\ee
for any measurable set $E \subset \mathcal M$.

Therefore a measure $\mu$ on a measurable space $\mathcal M$ is called
 $\mathcal G$$-$invariant, if
 \be
 \label{inv}
\mu(k^{-1}E)=\mu(E),
 \ee
for any $k \in \mathcal G$ and any $E \subset \mathcal M.$

The weaker condition of $\mathcal G$$-$quasi$-$invariant measure $\mu$, or measure
$\mu$ quasi$-$ \newline invariant with respect to $\mathcal G$, states that if $E$ has positive measure, then its image $k^{-1} E$ has also positive measure for any $k \in \mathcal G.$ Thus
\be
\label{con1}
\mu(E)>0 \Rightarrow \mu_{k}(E)>0,
\ee
for any $k \in \mathcal G.$
The last condition is equivalent to the statement that the measures $\mu$ and
$\mu_{k}$ have the same sets of measure zero, i.e.,
\be
\label{con2}
\mu(E)=0 \Rightarrow \mu_{k}(E)=0.
\ee

Quasi$-$invariance is necessary and sufficient for the existence of the Radon$-$\newline Nikodym (R$-$N) derivative
$\frac{\rm{d}\mu_{k}}{\rm{d}\mu}(p)$, $p \in \mathcal M$,
for all group elements $k \in \mathcal G.$ In the case
under consideration $\mathcal G= G = SL(2,R)$, and
$\mathcal M=G/L_{f}.$ Note that it is precisely the
quasi$-$invariance of $\mu$ with respect to $G$ that
allows the square root of the R$-$N derivative to occur as
a factor in (\ref{ukygvkhvkyutfkyufvmhgv}), and this is exactly what is needed to make
the representation of $\mathcal B$ in $H_{\mu}$,
given in (\ref{ukygvkhvkyutfkyufvmhgv}), unitary.

Construction of $\mathcal G$$-$quasi$-$invariant measures
on the
$\mathcal G$ orbits on the topological dual  $A^{\prime}$ of $A$  is a major mathematical problem in every group theoretical approach to quantum gravity which involves the
construction of the IRS of a symmetry group of the form $\mathcal B = A  \rtimes_{T} \mathcal G. $
In every approach this measure carries  important information.

When $\mathcal G$ is a not locally compact group
the construction of $\mathcal G$$-$quasi$-$invariant measures
on the
$\mathcal G$ orbits on the topological dual  $A^{\prime}$ of $A$ is an in general unsolved problem. This is the case
in Isham's approach (in Section \ref{other} we adopted a different notation; we denoted $\mathcal B$ by $\mathcal C$,
 $A$  by $\mathcal S$, and, $\mathcal G$ by $G$).
 In fact Isham conjectured that there are not any
$\mathcal G$$-$quasi$-$invariant measures
on the $\mathcal G$ orbits on the topological dual  $A^{\prime}$ in his approach and that it is likely that
such measures will  of necessity be concentrated
on genuine distributional elements in $A^{\prime}$.
If this is true, Isham concluded,
``the group representation theory
determines by itself precisely what is to be understood by `distributional geometry'  ''.
No matter how significant and insightful that is, it is still a conjecture.

The key fact in our approach which comes to our rescue is that
$\mathcal G=G=SL(2,R).$ $SL(2,R)$ is locally compact and therefore has a Haar measure. This is true for any
BMS group
$B=A \rtimes_{T} \mathcal G$, and for any of its analogues, in any number $d\geq3$
of space$-$time dimensions. Thus in all cases $\mathcal G$  is
locally compact.

If this had not been the case
we would have added as the fourth problem,
in Subsection \ref{extension}, the construction of
 $\mathcal G$$-$quasi$-$invariant measures
on the
$\mathcal G$ orbits on the topological dual  $A^{\prime}$,
to the three problems stated there, namely,
 the ``little groups'' problem, the irreducibility problem, and the
exhaustivity problem,
the solutions to which are necessary in order to
carry out the necessary extension of Wigner$-$Mackey's theory.

In our approach the  $\mathcal G$$-$quasi$-$invariant measures
on the
$\mathcal G$ orbits on the topological dual  $A^{\prime}$
carries the information about how the outcomes of physical
measurements distribute.

\subsubsection{The Haar measure}
The Haar measure is a generalization of the Lebesgue
measure to the case where the measurable space
$\mathcal M$ is a locally compact group $G$. The Haar measure
shares the key property of the Lebesgue
measure, namely it is invariant under translations.
Since the measurable space is a group it is natural to
distinguish left$-$invariance from right$-$invariance
since the translation is realized now by group multiplication.
In particular the following Theorem \cite{Haar} holds:
\begin{theorem}
Let $G$ be a locally compact group. Then there exists a unique, up to a positive multiplicative
constant, left$-$invariant measure, the Haar measure, on $G$. Similarly if
``left$-$invariant'' is replaced by``right$-$invariant''.
\end{theorem}

Let $ \Omega$ be the class of Borel sets in $G.$
A left invariant (right invariant) Haar measure on $G$
is a measure $\mu$ such that
\be
\label{la}
\mu(g^{-1}A)=\mu(A)
 \ee
($\mu(A g^{-1})=\mu(A)$), for all $g \in G$ and $A \in  \Omega.$ Note that according to the definition of
invariant measures given in equation (\ref{inv}) a left
invariant measure on $G$, is $G$$-$invariant.

Given a left Haar measure, one immediately obtains a right Haar measure, and
vice versa. In particular, if $\mu$  is a left (right) invariant
measure on $G$ then $\mu^{\prime}(A)=\mu(A^{-1})$
is a right (left) measure on $G$, where $A^{-1}$
contains the inverse elements of $A$.


In general, a left$-$invariant Haar measure on a locally compact group $G$, is not
right$-$invariant. When on a group $G$ the Haar measure is both left and right invariant
the group $G$ is called unimodular.

A subgroup of a locally compact group is locally compact. Therefore since $SL(2,R)$ is locally compact, its subgroups
$SO(2)$, and
$C_{n}$, where $n$ is even, are locally compact. As a matter of fact,  the groups $SO(2)$, and
$C_{n}$, where $n$ is even, are compact, and hence closed,
 subgroups of $SL(2,R)$. Compact groups are unimodular.
Therefore the groups
$SO(2)$, and
$C_{n}$, where $n$ is even, are unimodular.
Moreover, the group $SL(2,R)$ is also
unimodular \cite{Lang}.

The Haar measure on $SL(2,R)$ is given in \cite{Gel}, p. 214$-$215, the
Haar measure on $SO(2)$ is the usual
Lebesgue measure on $SO(2)$, and the Haar measure on $C_{n}$, with $n$ even, is the
counting measure on $C_{n}$ defined by:
\be
\mu: \ \mathcal P(C_{n}) \longrightarrow R_{\geq}, \quad \mu(E)=| E |,
\ee
where  $\mathcal P(C_{n})$ is the power set of $C_{n},$
$R_{\geq}$ is the set of non$-$negative real numbers, and $| E |$,
is the order of the set $E.$

\subsubsection{Projection of the Haar measure}

Consider the canonical map
\be
pr: \ G \longrightarrow G/L_{f},
\ee
where $G=SL(2,R)$, and $L_{f}=SO(2)$, or $L_{f}=C_{n}$, where $n$ is even.
It is a beautiful, intuitively  obvious, Theorem (\cite{Lang}, p. 37, Theorem 1)
that when $G$ and $L_{f}$ are unimodular, and $L_{f}$
is a closed subgroup of $G$,
the unique, up to an arbitrary multiplicative positive constant,
Haar measure on $G$  can be projected down uniquely, up to an arbitrary  multiplicative positive constant,
to a $G$$-$invariant  Haar measure on the coset space $G/L_{f}$, and hence
to a $G$$-$invariant  Haar measure on the orbit  $Gf.$

The Theorem asserts in particular that if $\mu_{G}$ and $\mu_{L_{f}}$
are the unique, up to an arbitrary multiplicative positive constant,
Haar measures on the groups $G$ and $L_{f}$ respectively, then there is
a unique, up to an arbitrary multiplicative positive constant,
  $G$$-$invariant  Haar measure $\mu_{G/L_{f}}$ on the coset space $G/L_{f}.$
 Thus there is one$-$to$-$one correspondence
 \be
(\mu_{G},\mu_{L_{f}}) \longleftrightarrow \mu_{G/L_{f}},
 \ee
between the pairs of measures $(\mu_{G},\mu_{L_{f}})$  and the measures $\mu_{G/L_{f}}$ on the coset spaces $G/L_{f}.$
Moreover, the Theorem gives explicitly $\mu_{G/L_{f}}$, and it describes
 how $\mu_{G/L_{f}}$ is constructed from the measures $\mu_{G}$ and $\mu_{L_{f}}.$


We note that the projection from the Haar measure on $G$ to the
$G$$-$invariant  Haar measure on the orbit $Gf \simeq G/L_{f}$ is
obvious when $G$ and $L_{f}$ are finite groups and the Theorem in question
generalizes beautifully this  obvious fact.
The sought
$G$$-$invariant measures on the orbits $Gf \simeq G/L_{f}$,
where $G=SL(2,R)$, and $L_{f}=SO(2)$, or $L_{f}=C_{n}$, where $n$ is even,
are obtained immediately by  applying this Theorem.
Details are given in \cite{Melas}.

This completes the necessary information in order to construct the induced representations of $B(2,1)$.
As already stated, it is shown in \cite{Melas} that
when $B(2,1)$ is equipped with the Hilbert topology the inducing construction is exhaustive notwithstanding the fact that $B(2,1)$ is not locally compact.
 To conclude, in this paper \it all \normalfont IRS
of $B(2, 1)$ have been constructed in the Hilbert topology.

%




\end{document}